\documentclass[a4paper, 11pt]{article}
\usepackage{aapreprint}
\usepackage{amsmath}
\usepackage{appendix}
\usepackage{graphicx}
\usepackage{subfigure}
\usepackage{amssymb}
\usepackage{xcolor}
\usepackage{multirow}
\usepackage{cancel}
\usepackage{color}
\usepackage{ulem}
\usepackage{caption}
\usepackage{listings}
\usepackage{xcolor}
\usepackage{float}
\usepackage{slashed}

\usepackage{graphicx}

\usepackage{float}
\usepackage{amsmath}
\usepackage{amssymb}
\usepackage[utf8]{inputenc}
\usepackage{hyperref}
\usepackage{color}
\usepackage{slashed}
\usepackage[absolute]{textpos}
\usepackage{changes}
\allowdisplaybreaks

\graphicspath{{images/}}

\newcommand{\be}{\begin{equation}}
\newcommand{\ee}{\end{equation}}

\begin{document}
\title{
Disentangling Dark Gauge Symmetries with Deep Learning on the Lund jet plane}
\author[1]{Jinmian Li,}
\emailAdd{jmli@scu.edu.cn}
\affiliation[1]{College of Physics, Sichuan University, Chengdu 610065, China}

\author[2]{Junle Pei,}
\emailAdd{peijunle@hnas.ac.cn}
\affiliation[2]{Institute of Physics, Henan Academy of Sciences, Zhengzhou 450046, China}

\author[1]{Rao Zhang}
\emailAdd{rzhang9527@gmail.com}

\abstract{ 

While dark sectors with new confining gauge symmetries are compelling candidates for resolving the dark matter mystery, discerning the underlying dark gauge group structure remains a significant phenomenological challenge. 
In this work, we systematically investigate the distinct radiation patterns of dark quarks and gluons by developing a novel Monte Carlo parton shower simulation framework applicable to arbitrary gauge groups. 
To handle generalized color topologies and the momentum recoil scheme, our algorithm constructs color dipoles using a group-theoretic tagging procedure. 
Furthermore, our simulation framework employs an exact three-body phase-space parameterization by analytically solving the cubic kinematic equation for each branching. 
This enables capturing full mass effects for both dark quarks and dark gluons, naturally yielding precise boundaries including mass-induced gaps and dead-cone thresholds. 
To decode these complex emission topologies, we utilize the Lund Jet Plane representation alongside a dedicated Neural Sorter Mamba Network. 
We demonstrate that our framework can successfully disentangle the perturbative footprints of different gauge symmetries. Finally, we show that our discrimination power remains robust against the unknown non-perturbative details of dark hadronization by maintaining high classification efficiencies even under stringent infrared $k_T$ cutoffs, and we explicitly quantify the impact of massive dark gauge bosons on the classification sensitivity.

}

\maketitle

\section{Introduction}

Although the Standard Model (SM) of particle physics has achieved remarkable success, it leaves several fundamental questions unanswered, most notably the nature of Dark Matter (DM) and the origin of the electroweak hierarchy~\cite{Gildener:1976ai}. A compelling class of solutions postulates the existence of a ``Hidden Valley'' or a ``Dark Sector''~\cite{Strassler:2006im,Han:2007ae,Gori:2022vri}, containing new degrees of freedom that are neutral under the SM gauge group but charged under a new confining gauge symmetry. Such scenarios naturally arise in models of Neutral Naturalness~\cite{Chacko:2005pe, Burdman:2006tz, Burdman:2008ek,Cai:2008au}, Composite Higgs~\cite{Weinberg:1975gm,Susskind:1978ms,Cacciapaglia:2020kgq}, and various Strongly Interacting Massive Particle (SIMP) dark matter candidates~\cite{Hochberg:2014dra,Hochberg:2014kqa,Lee:2015gsa,Hochberg:2015vrg}.

The phenomenology of these confining dark sectors is rich and distinct. If the confinement scale is low compared to the production energy scale, dark partons produced at high-energy colliders will undergo dark parton showering and hadronization, resulting in a collimated spray of dark hadrons~\cite{Cohen:2017pzm,Cohen:2020afv,Knapen:2021eip,Albouy:2022cin}. Depending on the lifetimes and decay modes of the resulting dark hadrons, these showers can manifest as emerging jets~\cite{Schwaller:2015gea,CMS:2018bvr,Carrasco:2023loy,ATLAS:2025lfx,Carrasco:2025bct}, semi-visible jets~\cite{Cohen:2015toa,Beauchesne:2017yhh,CMS:2021dzg,Liu:2024rbe,Buckley:2025hty,ATLAS:2025kuz}, displaced signal~\cite{Cheng:2021kjg,Born:2023vll,Cheng:2024hvq,Cheng:2024aco,Liebersbach:2024kzc,Liu:2025bbc,CMS:2025fnr,Chen:2025btv,Borrello:2026bfr}, soft unclustered energy patterns (SUEPs)~\cite{Harnik:2008ax,Knapen:2016hky,CMS:2024nca}, or stopping dark mesons~\cite{Asadi:2025vfr}. Beyond collider signatures, stable mesons/baryons in the dark sector are prime candidates for DM within the SIMP scenario\cite{Tsai:2020vpi,Contino:2020god,Kondo:2022lgg,Braat:2023fhn,Fleming:2024flc,Garcia-Cely:2024ivo,Asadi:2024tpu,Carmona:2024tkg,Frumkin:2025iit,Garcia-Cely:2025flv,Alfano:2025non}, naturally evading stringent direct detection constraints while potentially resolving the small-scale structure problem~\cite{Bullock:2017xww,Tulin:2017ara}. 

While significant efforts have been dedicated to searching for dark parton shower signatures, most studies predominantly assume the dark gauge group to be $SU(N)$, mirroring the QCD structure of the Standard Model. Simulations of such dark showers are typically implemented within the Pythia8~\cite{Carloni:2010tw,Carloni:2011kk,Bierlich:2022pfr} or Herwig7~\cite{Kulkarni:2024okx,Kim:2026mfk} frameworks, while the GlueShower package~\cite{Curtin:2022tou} addresses the pure Yang-Mills glueball case. However, from a theoretical perspective, the dark gauge group is largely unconstrained and could plausibly be based on other Lie groups, such as the orthogonal $SO(N)$ or symplectic $Sp(2N)$ groups. Recent theoretical works have systematically examined dark matter candidates under these various canonical group structures. By combining effective field theory with lattice field theory calculations, detailed analyses of self-interaction characteristics have been conducted for $SO(N)$~\cite{Pomper:2024otb} and $Sp(2N)$~\cite{Kulkarni:2022bvh,Zierler:2022uez,Dengler:2024maq,Dengler:2025ulb,Kolesova:2025ghl} theories, respectively. Despite these advances in understanding the low-energy spectrum, the high-energy radiation patterns unique to these groups remain underexplored. 

Since the group structure fundamentally dictates the Casimir invariants and color factors governing parton evolution, distinguishing the underlying gauge symmetry is crucial for reconstructing the ultraviolet (UV) nature of the dark sector. Discerning the gauge structure of a secluded sector is, however, experimentally challenging. Previous works have explored the distinction between Abelian $U(1)$ and non-Abelian $SU(N)$ scenarios using event shapes and lepton spectra~\cite{Carloni:2011kk}. Discriminating between different non-Abelian groups (e.g., $SO(N)$ vs. $Sp(2N)$) is far more intricate, requiring probes sensitive to subtle differences in radiation patterns within the dark shower. This task is further complicated by the unknown nature of dark hadronization. As the non-perturbative dynamics of general gauge groups are not well understood from first principles, it is imperative to identify observables that are robust against hadronization uncertainties and sensitive primarily to the perturbative parton shower evolution.

In this work, we systematically investigate the radiation patterns of dark quarks and gluons under $SU(N)$, $Sp(2N)$, and $SO(N)$ gauge symmetries. 
We have developed a Monte Carlo simulation framework that implements parton showers for these general gauge groups. 
We construct color dipoles using a group-theoretic tagging scheme. Each parton carries a set of discrete gauge-group labels, which encode its color connections in a representation-independent way. A dipole forms between two partons when their tags match within conjugate representation, generalizing the standard QCD color-flow picture to arbitrary gauge groups and representations. 
This tagging mechanism also identifies the recoil partner for each branching. The parton sharing a color line with the splitting parton absorbs the longitudinal recoil, ensuring local four-momentum conservation without ad hoc prescriptions. This formalism applies directly to both massless and massive partons, as well as mixed-representation systems (such as an adjoint particle coupling to two fundamental-anti-fundamental dipoles).

Including full mass effects for both dark quarks and dark gluons is important for realistic modeling. 
Finite masses suppress collinear radiation via the ``dead cone'' effect~\cite{Dokshitzer:1991fd} at emission angles below $m/E$. They also introduce significant corrections in the ultra-collinear regime~\cite{Chen:2016wkt,Dittmaier:2025htf}, modifying the splitting kernels at ${\cal O}(m^2/p_T^2)$. 
To accurately model these kinematics, we employ an exact three-body phase-space parameterization. Rather than relying on massless approximations, we determine the allowed ($t$, $z$) region for each branching by analytically solving the cubic kinematic equation. This yields precise boundaries, including mass-induced gaps and dead-cone thresholds. 
The corresponding splitting functions similarly incorporate complete mass corrections, interpolating between the massless DGLAP kernels and the ultra-collinear limit where mass terms dominate. 
Since massive partons are common in BSM scenarios, capturing these phase-space modifications is necessary for reliable predictions of dark-sector phenomenology at colliders and in indirect detection experiments.

To extract the maximum amount of information from the complex final states, we utilize the Lund Jet Plane (LJP)~\cite{Dreyer:2018nbf} representation. The LJP provides a powerful visualization of the phase space of jet fragmentation, effectively separating perturbative radiation from non-perturbative hadronization effects~\cite{Dreyer:2020brq,Dreyer:2021hhr,Cohen:2023mya,Baldenegro:2024pfb}. By treating the LJP representations as sequences, we employ Neural Sorter Mamba Network (NS-MambaNet) to discriminate between jets originating from different gauge groups. Furthermore, to demonstrate the robustness of our discrimination power against the unknown details of dark hadronization, we analyze the performance under varying cuts on the relative transverse momentum $k_T$ of the emissions. By imposing a $k_T$ cut, we effectively remove the infrared soft emissions dominated by non-perturbative physics, thereby isolating the perturbative footprints determined by the gauge structure.

This paper is organized as follows. In Section~\ref{sec:2}, we provide the details of the model setup and calculate the splitting function in each case. In Section~\ref{sec:3}, we review the theoretical framework for parton showers in general gauge groups and describe our simulation setup, including the treatment of massive partons. Section~\ref{sec:4} details the construction of the Lund Jet Plane images and the architecture of the neural networks used for classification. In Section~\ref{sec:5}, we present the discrimination results for various model benchmarks and discuss the impact of $k_T$ cuts on the classification performance. We summarize our findings in Section~\ref{sec:6}. 
We also provide Appendix~\ref{sec:app1} and~\ref{sec:app2} to detail the mathematical derivations of the kinematically allowed phase space boundaries for massive splittings, and to compare the kinematic distributions generated by our algorithm with those from Pythia8. 

\section{The Lagrangian for dark sector}\label{sec:2}

To correspond with the QCD of the SM, we propose the existence of a non-Abelian gauge interaction, including \(SU(N)\), \(Sp(2N)\), \(SO(N)\), in the dark sector, which exhibits asymptotic freedom and dark quark confinement. The Lagrangian describing dark quarks and dark gluons can be written as
\begin{align}
    \mathcal{L}_{\mathrm{dQCD}} = -\frac{1}{4} \sum_{a=1}^{n_V} G_{\mu \nu}^a G_a^{\mu \nu} + \sum_{k,j=1}^{N_f} \bar{q}_k(i\slashed{D}\delta_{k,j} - m_{k,j}) q_j~,\label{lag}
\end{align}
where \( G_{\mu \nu}^a \) (\( a = 1, 2, \ldots, n_V \)) denotes the field-strength tensor of the dark gluons, and $n_V$ is given by $N^2-1$, $N(2N+1)$, and $N(N-1)/2$ for the gauge groups \(SU(N)\), \(Sp(2N)\), and \(SO(N)\), respectively. The term \( m_{k,j}~(k,j=1,2,\ldots,N_f) \) denotes the elements of the mass matrix \( M_q \) for the dark quarks \( q_k \) (\( k = 1, 2, \ldots, N_f \)). 

For generality, we also consider the case where the dark gauge boson is massive, for instance due to a dark Higgs mechanism or other effects. In such a case, the gauge boson acquires an additional longitudinal polarization, whose splitting kinematics differ dramatically from those of the transverse components. These differences leave imprints on jet substructure variables, such as the Lund jet observables. However, the specific details of the spontaneous gauge symmetry breaking lie beyond the scope of the present study. For simplicity, we will assign a universal mass to all gauge bosons within a given model.

\subsection{The Splitting fucntion}

When many external legs are involved, computing amplitudes and cross sections with the conventional Feynman-diagram approach becomes impractical. A standard workaround is to employ the parton-shower framework \cite{Collins:1989bt,Hoche:2014rga,Nagy:2017ggp,Forshaw:2020wrq,Papaefstathiou:2024qlg}, which breaks the full cascade into successive $1 \rightarrow 2$ splittings. The differential cross section for a hard process followed by the branching $A \to B + C$ can be written in factorized form as
\begin{align}
  d \sigma_{X,BC}
\simeq
  d \sigma_{X,A}\times  d \mathcal{P}_{A \rightarrow B+C}~,
\end{align}
where $X$ denotes the remaining particles produced in the hard interaction, excluding $A$. The quantity $d \mathcal{P}_{A \rightarrow B+C}$ is the differential splitting probability for the transition $A \to B + C$,
\begin{align}
  \frac{d \mathcal{P}_{A \rightarrow B+C}}{d z ~d \ln Q^{2}}
\approx
  \frac{1}{S}
  \frac{1}{16 \pi^{2}}
  \frac{Q^2}{\left(Q^{2}-m_{A}^{2}\right)^{2}}
  \left|\mathcal M_{\text{split}}\right|^{2},
\label{split}
\end{align}
with $z$ the energy fraction carried by $B$ and $Q^2$ the virtuality of the propagating $A$. The squared matrix element ${\left|\mathcal M_{\text{split}}\right|^2}$ is obtained from the amputated diagram for $A \to B + C$, evaluated with on-shell polarization vectors. The symmetry factor $S$ is 2 when $B$ and $C$ are identical and 1 otherwise.

For a time-like splitting $A \rightarrow B + C$, we parametrize the particle momenta as
\begin{subequations}
\begin{align}
& P_A \equiv \left(E_A, 0, 0, E_A - \frac{k_T^2+\bar{z}m_B^2+zm_C^2}{2z\bar{z}E_A}\right), \\
&P_B \equiv \left(zE_A, k_T, 0, zE_A - \frac{k_T^2 + m_B^2}{2z E_A}\right), \\
&P_C \equiv \left(\bar{z}E_A, -k_T, 0, \bar{z}E_A - \frac{k_T^2 + m_C^2}{2\bar{z}E_A}\right),
\end{align}
\end{subequations}
where the energy fractions $z$ and $\bar{z} \equiv 1 - z$ lie within $(0, 1)$. We assume $E_A^2$ is much larger than the transverse momentum $k_T^2$ and the masses $m_i^2$ for $i = A, B, C$. Discarding terms of order $(k_T^2 \text{ or } m_i^2) / E_A^2$ for $i = A, B, C$, the corresponding virtualities are
\begin{align}
P_A^2 \equiv Q^2=\frac{k_T^2+\bar{z}m_B^2+z m_C^2}{z\bar{z}},
\quad
P_B^2 = m_B^2,
\quad
P_C^2 = m_C^2.
\end{align}
Particles $B$ and $C$ are on shell, whereas $A$ carries virtuality $Q$.

\begin{table}[htb]
	\centering
	\begin{tabular}{c|c}  
		$A\rightarrow B+C$&   $\frac{d \mathcal{P}_{A \rightarrow B+C}}{d z~ d\ln Q^{2}}=P_{A \rightarrow B+C}(z)$\\
		\hline 
		$V_{T} \rightarrow \bar{q} / q+q / \bar{q}$ & $\frac{\alpha}{2 \pi}\frac{1}{Q^2}  \left(Q^2\left(z^{2}+\bar{z}^{2}\right)+2m_q^2\right)  T_R$\\
        $q / \bar{q} \rightarrow V_{T}+q / \bar{q}$ &   $\frac{\alpha}{2 \pi}\frac{Q^2}{\left(Q^{2}-m_{q}^{2}\right)^2}  \left(Q^2\frac{1+\bar{z}^{2}}{z}- m_{q}^{2}\frac{2+z^2}{z}\right)  C_F$\\
        $V_T \to V_T+V_T$ &  $\frac{\alpha}{2\pi}\frac{\left(1-z+z^2\right)^2}{ z(1-z) }  C_A$
	\end{tabular}
	\caption{\label{tab:splitf-massless} Splitting functions involving $V_T$ and $q/\bar{q}$ with unbroken gauge symmetry.} 
\end{table}

For the dark gluon ($V$)–dark quark ($q$) interactions encoded in Eq. (\ref{lag}), and in the unbroken gauge-symmetry case (so that the gauge boson $V$ is massless), the splitting functions for time-like branchings are collected in Table \ref{tab:splitf-massless}. All results are averaged over the polarizations of the incoming particles and summed over those of the final states.
The dark fine-structure constant $\alpha$ is defined as $\alpha \equiv g_D^2 / 4\pi$, and $m_q$ is the dark quark mass.
The coefficients $T_R$, $C_F$, and $C_A$ are fixed by the gauge group ($G$) and by the representation ($r$) of the dark quark. Assuming the dark quark transforms in the fundamental representation of $G$, we denote the generators in this representation by $t^a_{m,n}$ and the structure constants by $f^{abc}$. Here $a=1,2,...,n_V$ labels the generators, and hence the dark-gluon colors, while $m,n=1,2,...,n_r$ run over the $n_r$ colors of the dark quark. 
In the fundamental representation, $n_r$ is $N$ for the gauge groups \(SU(N)\) and \(SO(N)\), and $2N$ for the gauge group \(Sp(2N)\).
With these conventions, we have
\begin{align}
  &  T_R=\frac{1}{n_V}\sum_{a=1}^{n_V}\text{Tr}[t^at^a]~, \\
   &  C_F=\frac{1}{n_r}\sum_{a=1}^{n_V}\text{Tr}[t^at^a]~, \\
    &  C_A=\frac{1}{n_V}\sum_{a,b,c=1}^{n_V}f^{abc}f^{abc}~.
\end{align}
Therefore, $T_R$ and $C_F$ denote the Dynkin index and the quadratic Casimir of the fundamental representation, respectively, while $C_A$ denotes the quadratic Casimir of the adjoint representation (which is also the Dynkin index of the adjoint representation).
For $G=U(1)$, $SU(N)$, $Sp(2N)$, and $SO(N)$, the corresponding values of $T_R$, $C_F$, and $C_A$ are provided in Table \ref{tab:splitf-coe}. In the $U(1)$ case, $\mathcal{Q}$ denotes the dark charge carried by the dark quark.

\begin{table}[htb]
	\centering
	\begin{tabular}{c|ccc}  
		 Gauge symmetry &   $T_R$ &    $C_F$  &   $C_A$ \\
		\hline 
		$U(1)$ &  $\mathcal{Q}^2$  & $\mathcal{Q}^2$ & $0$ \\
        $SU(N)$ & $\frac{1}{2}$ & $\frac{N^2-1}{2N}$ & $N$ \\
        $Sp(2N)$ & $\frac{1}{2}$ & $\frac{2N+1}{4}$ & $N+1$ \\
        $SO(N)$ & $1$ & $\frac{N-1}{2}$ & $N-2$ \\
        \end{tabular}
	\caption{\label{tab:splitf-coe} $T_R$, $C_F$, and $C_A$ under different gauge symmetries.} 
\end{table}

When the gauge symmetry $G$ is broken, for simplicity we take all color states of the gauge boson (dark quark) to share a common mass $m_V$ ($m_q$). The gauge boson $V$ then carries both transverse modes $V_T$ and a longitudinal mode $V_L$. Within this simplified setup, splitting functions that include gauge-boson mass effects are presented in Table \ref{tab:splitf-massive}. In particular, for $V\to V+V$ we sum over all polarization states of each daughter $V$ and average over the polarizations of the parent $V$.
When computing splitting functions involving the longitudinal mode of the dark gluon, it is essential to discard terms proportional to $Q^2-m_A^2$ \cite{Chen:2016wkt}.
The function for the process $q / \bar{q} \rightarrow q / \bar{q} + V_{T/L}$ follows from that for $q / \bar{q} \rightarrow V_{T/L} + q / \bar{q}$ via
$ P_{A\rightarrow B+C}(z) = P_{A\rightarrow C+B}(\bar{z})$.

\begin{table}[htb]
	\centering
	\begin{tabular}{c|c}  
		$A\rightarrow B+C$&   $\frac{d \mathcal{P}_{A \rightarrow B+C}}{d z~ d\ln Q^{2}}=P_{A \rightarrow B+C}(z)$\\
		\hline 
		$V_{T} \rightarrow \bar{q} / q+q / \bar{q}$ & $\frac{\alpha}{2 \pi}\frac{Q^2}{\left(Q^{2}-m_{V}^{2}\right)^2}  \left(Q^2\left(z^{2}+\bar{z}^{2}\right)+2m_q^2\right)  T_R$\\
        $V_{L} \rightarrow \bar{q} / q+q / \bar{q}$  & $\frac{2 \alpha}{\pi}  \frac{Q^2}{\left(Q^2-m_{V}^{2}\right)^{2}}m_{V}^{2} z \bar{z} T_R$\\
        $q / \bar{q} \rightarrow V_{T}+q / \bar{q}$ &   $\frac{\alpha}{2 \pi}\frac{Q^2}{\left(Q^{2}-m_{q}^{2}\right)^2}  \left(Q^2\frac{1+\bar{z}^{2}}{z}- m_{q}^{2}\frac{2+z^2}{z}-m_{V}^{2}\frac{1+\bar{z}^{2}}{z^2}\right)  C_F$\\
        $q / \bar{q} \rightarrow V_{L}+q / \bar{q}$ & $\frac{\alpha}{\pi}  \frac{Q^2 }{\left(Q^{2}-m_{q}^{2}\right)^{2}}m_{V}^{2} \frac{\bar{z}}{z^2}  C_F$\\
        $V \to V+V$ &  $\frac{\alpha}{2\pi}\frac{Q^2}{\left(Q^2-m_{V}^2\right)^2}\frac{9  \left(4 Q^2 \left(1-z+z^2\right)^2-m_{V}^2 \left(4 z^4-8 z^3+z^2+3 z+4\right)\right)}{ 32z(1-z) }  C_A$
	\end{tabular}
	\caption{\label{tab:splitf-massive} Splitting functions involving $V_{T/L}$ and $q/\bar{q}$ with broken gauge symmetry.} 
\end{table}

\subsection{Running Couplings in the Dark Sector}
The dynamics of the dark parton shower are intimately related to the behavior of the strong coupling in the dark sector $\alpha$. Similar to the QCD, the dark gauge coupling $g_D$ is not a constant but evolves with the energy scale $\mu$. This scale dependence determines the intensity of perturbative dark parton emissions and dictates the energy scale at which the theory becomes strongly coupled. At the one-loop level, the evolution of the gauge coupling is governed by the Renormalization Group Equation (RGE):
\begin{align}
\frac{dg_D}{d \ln \mu} \equiv \beta_g = - \frac{B}{16 \pi^2} g_D^3
\end{align}
where $B$ is the coefficient of the $\beta$-function. By integrating this differential equation, we obtain the explicit scale dependence of the running coupling:
\begin{align}
\alpha(\mu) = \frac{2 \pi}{B \ln (\mu/\Lambda)} ~,~ g_D^2(\mu)=\frac{8 \pi^2}{B \ln ({\mu}/{\Lambda})}~.~
\end{align}
Here, $\alpha(\mu) \equiv g_D^2(\mu)/(4\pi)$, and $\Lambda$ represents the intrinsic confinement scale of the dark sector (analogous to $\Lambda_{\text{QCD}}$), at which the perturbative coupling diverges and non-perturbative hadronization effects take over. In this work, without loss of generality, we set $\Lambda=0.1$ GeV. 

For a general non-Abelian gauge theory with a gauge group $G$ and fermionic matter $f$, the one-loop $\beta$-function is expressed as:
\begin{align}
\beta_g = -\frac{g_D^3}{16 \pi^2} [ \frac{11}{3} C_A - \frac{4}{3} \sum_f \kappa T_R(f) ] ~,
\end{align}
where the first term arises from the gauge boson self-interactions, and the second term is attributed to fermion vacuum polarization. The factor $\kappa=1~(1/2)$ for Dirac (Weyl) fermions. 

Assuming the presence of $N_f$ flavors of Dirac fermions ($\kappa=1$) in the fundamental representation, the fermion contribution to the $\beta$-function is multiplied by $N_f$. Depending on the choice of the underlying dark gauge group, the Dynkin index and the quadratic Casimir take specific values, leading to distinct running behaviors:
\begin{itemize}
\item $SU(N)$ gauge group, $C_A=N$, $T_R=1/2$,
\begin{align}
\beta_g=-\frac{g_D^3}{16 \pi^2}[\frac{11}{3} N - \frac{2}{3} N_f]
\end{align}
\item $Sp(2N)$ gauge group, $C_A = N+1$, $T_R=1/2$,
\begin{align}
\beta_g=-\frac{g_D^3}{16 \pi^2} [\frac{11}{3} (N+1) - \frac{2}{3} N_f]
\end{align}
\item $SO(N)$ gauge group, $C_A = N-2$, $T_R=1$,
\begin{align}
\beta_g = - \frac{g_D^3}{16 \pi^2} [\frac{11}{3} (N-2) - \frac{4}{3} N_f]
\end{align}
\end{itemize}

For the dark sector to exhibit asymptotic freedom, the overall $\beta$-function coefficient $B$ must strictly remain positive. This property ensures that the theory is weakly coupled at high energies, allowing initial hard partons to initiate a perturbative shower that manifests as collimated jet-like structures rather than a spherical spray of soft particles. Imposing $B > 0$ yields a theoretical upper bound on the number of dark quark flavors, $N_f$, which depends on the specific choice of the gauge group and fermion representation~\cite{Kulkarni:2025rsl}. 

\begin{figure}[htbp]
\includegraphics[width=0.3\textwidth]{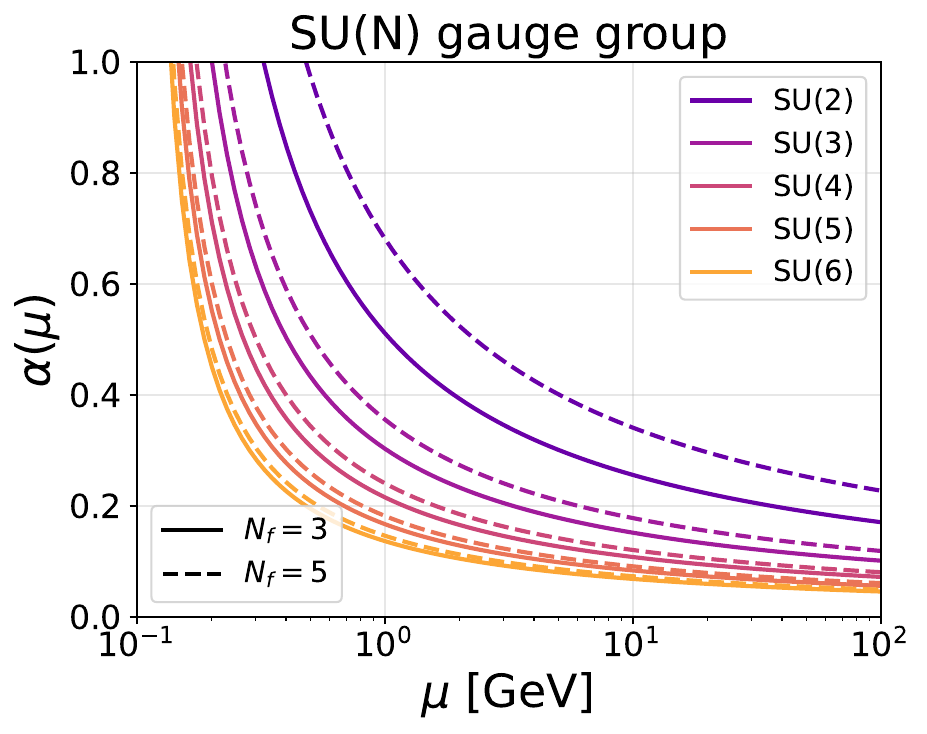}
\includegraphics[width=0.3\textwidth]{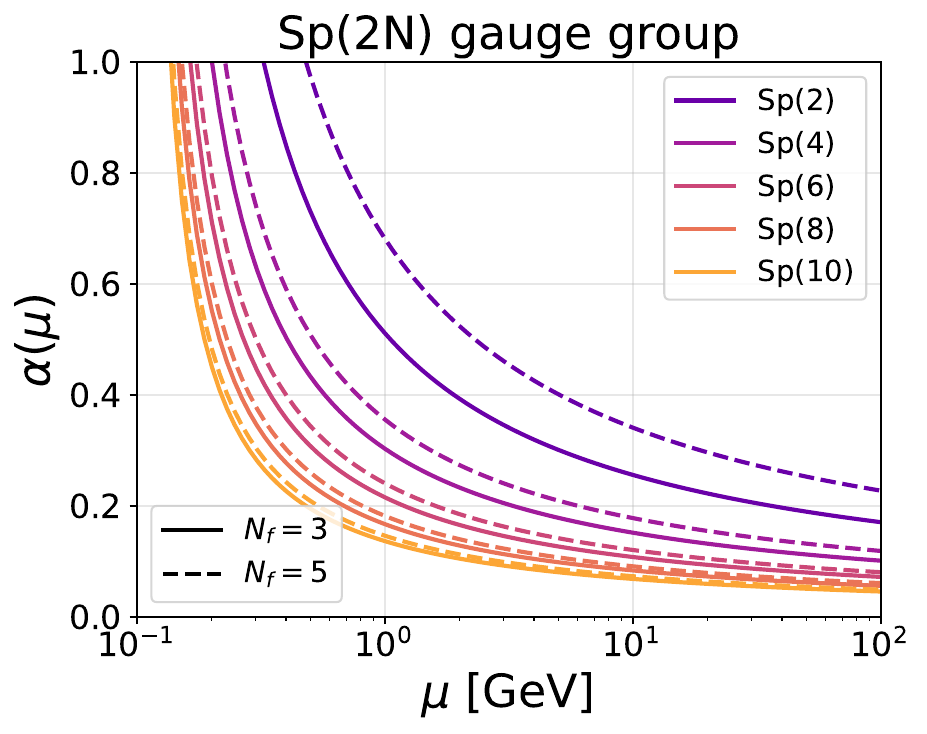}
\includegraphics[width=0.3\textwidth]{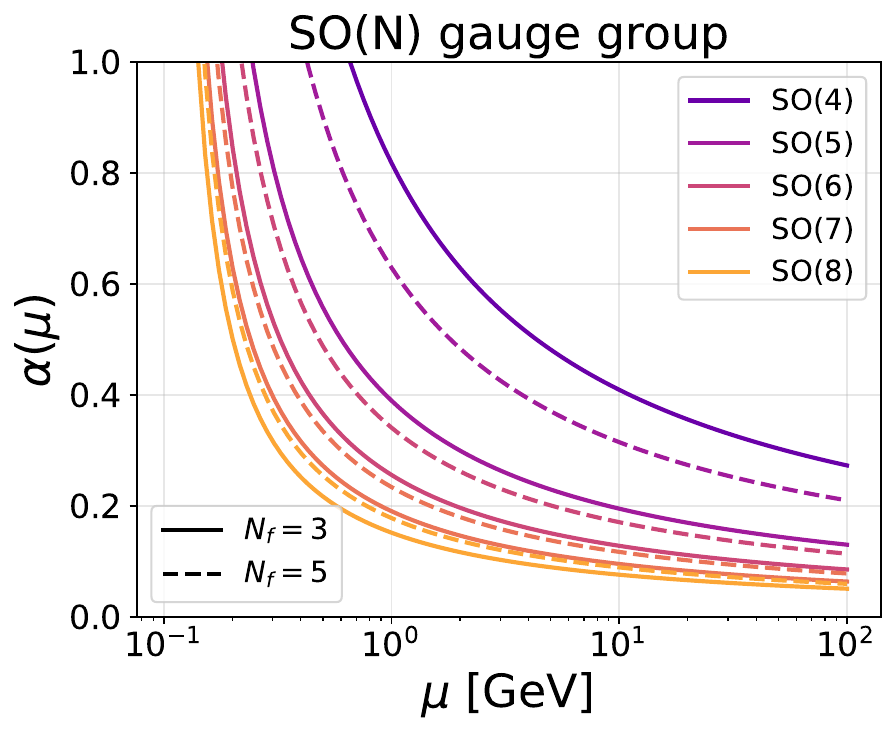}
\caption{\label{fig:ps} The RGE running of the gauge couplings $\alpha(\mu)$ as a function of the energy scale $\mu$ for $SU(N)$, $Sp(2N)$, and $SO(N)$ gauge groups. The variations in the slopes explicitly demonstrate how the choice of the gauge group and $N_f$ dictates the running behavior of the dark sector.}
\end{figure}

As illustrated in Figure~\ref{fig:ps}, the evolution trajectory of $\alpha(\mu)$ varies significantly across different gauge groups. Groups with larger Casimir invariants, such as higher-rank $SU(N)$ or $Sp(2N)$ groups, possess larger $\beta$-function coefficients, leading to a steeper running of the coupling. Consequently, even if one assumes a fixed confinement scale $\Lambda$ across different models, the effective coupling $\alpha(k_T)$ evaluated at the characteristic shower evolution scale (e.g., the transverse momentum $k_T$ of the splitting) will differ substantially depending on the underlying gauge symmetry. 

This differential running behavior is critical for the phenomenology of dark sector showers. Since the Altarelli-Parisi splitting kernels are directly proportional to the running coupling evaluated at the splitting scale, a steeper running coupling dynamically enhances soft and collinear radiation relative to a theory with a flatter running coupling. This scale-dependent emission probabilities leaves distinct, measurable imprints on the internal jet substructure. Furthermore, in a quark jet, the relative rates of primary quark emissions versus secondary gluon branchings are highly gauge-dependent. These branching rates are fundamentally dictated by the distinct ratios of the fundamental and adjoint Casimir invariants ($C_F$ versus $C_A$). Together with the running of the coupling, these structural features are deeply encoded in the multi-dimensional phase space of the shower. Ultimately, these intertwined perturbative signatures provide the robust physical foundation for discriminating between different dark gauge symmetries.

\section{Framework of the Monte Carlo Simulation of parton shower} \label{sec:3}
The simulation of parton showers in the standard approach is modeled as Markov Chain Monte Carlo (MCMC) processes based on either collinear Altarelli-Parisi splitting functions or the dipole/antenna formalism. These classical methods rely on probabilistic approximations, enforcing momentum conservation and color coherence (e.g., angular ordering) at each splitting step. More recently, there has been a surge of interest in simulating parton showers using quantum computing algorithms~\cite{Yamazaki:2025bfq}. Instead of relying on classical probabilities, these quantum approaches evolve the state vector directly. This formulation allows them to natively capture quantum interference effects~\cite{Bauer:2019qxa, Gustafson:2022dsq,Deliyannis:2022uyh}, explore novel algorithmic paradigms such as quantum walks to efficiently sample emission histories~\cite{Bepari:2021kwv}, investigate specific phenomenological targets like dark sector radiation~\cite{Chigusa:2022act}, and rigorously track branching kinematics within scalable hybrid frameworks~\cite{Bauer:2023ujy}. However, while quantum algorithms represent a promising frontier, they are currently limited by hardware constraints. 
Existing classical general-purpose event generators often require complex modifications to accommodate generalized Dark Sector topologies, particularly when non-canonical gauge groups or arbitrary mass spectra are introduced.

In this section, we present a generalized numerical Monte Carlo algorithm designed to simulate $p_T$-ordered dipole showers for arbitrary gauge symmetries, including $U(1)$, $SU(N)$, $SO(N)$, and $Sp(N)$. The core architecture and physical framework are summarized as follows:

\begin{enumerate}
\item \textbf{Generic Gauge Group:}
The algorithm is formulated within a generalized Lie algebra framework, supporting arbitrary gauge groups and their associated representations. By evaluating group-theoretic factors (e.g., quadratic Casimir operators $C_F/C_A$ and Dynkin indices $T_R$) based on particle representation assignments, the algorithm is universally applicable to the SM and various BSM scenarios involving $SO(N)$ or $Sp(N)$ sectors. 

The treatment of color connections is dynamically adapted to the underlying symmetry. Specifically, the algorithm constructs dipoles by pairing emitters with spectators that share identical gauge charge tags to reconstruct the charge flow topology. If no matching tags exist, the pairing falls back to the spectator that minimizes a charge-weighted kinematic distance defined as $d^2 / Q_r^2$. Here, $d^2 = (p_a + p_r)^2 - (m_a + m_r)^2$ represents the Lorentz-invariant interval between the emitter $a$ and spectator $r$, and $Q_r$ is the gauge charge of the spectator. Furthermore, for particles carrying multiple gauge charges, the framework allows a single particle to simultaneously evaluate emissions across all of its active gauge interactions, with a unified competition subsequently selecting only the highest-scale splitting to proceed. This structure also integrates the specific RGE running of the gauge couplings discussed in Section~\ref{sec:2}, evaluating the coupling dynamically at the local splitting scale.

\item \textbf{Evolution Dynamics:}
The evolution is governed by $p_T$ ordering, where emissions are generated in descending order of transverse momentum, modeling the leading-logarithmic behavior of the emissions. 
To rigorously incorporate mass effects, the mapping between the evolution scale $t$ (representing the relative transverse momentum squared, $k_T^2$), the light-cone momentum fraction $z$, and the virtuality $Q^2$ follows the exact kinematic relation:
\begin{align}
Q^2 = \frac{t + (1-z)m_B^2 + z m_C^2}{z(1-z)}~.~
\end{align}
This ensures that the phase space boundaries are strictly respected during the evolution, translating massive kinematics into the suppression of strictly collinear radiation known as the ``dead-cone'' effect.
The evolution proceeds via logarithmic scale discretization. To accurately treat particles with multiple indices of the same gauge group (e.g., gluons) while precluding double counting, the framework aggregates all active dipoles associated with a multi-indexed emitter to determine a unified trial emission probability: (1) Component dipoles connected to different spectators within the same gauge group are summed to obtain the total integral width: $\mathcal{I}_{\text{total}} = \sum_i \mathcal{I}_i$; (2) The probability of non-emission between two scales is governed by the Sudakov form factor. Within a discrete step $\Delta \ln t$, the emission probability is defined by $P_{\text{emit}} \approx (1 - e^{-\Delta \ln t}) \times \mathcal{I}_{\text{total}}$. If an emission is triggered, the specific radiative channel and splitting variables are sampled based on the relative weights of the individual dipole components. At the global level, every radiator makes an independent emission decision based on its own Sudakov factor within each discrete step. If multiple radiators produce trial emissions in the same step, only the candidate with the largest $t$ is retained and processed; the others are discarded and reconsidered in subsequent steps. The monotonic decrease of the evolution scale across steps ensures that all accepted emissions are strictly ordered in $p_T$.

\item \textbf{Kinematic Reconstruction and Local Recoil Scheme:} The algorithm implements a local recoil scheme to reconstruct the final-state momenta for the $2 \to 3$ splitting process $a + r \to b + c + r'$, where $a$ and $r$ denote the emitter and the spectator, respectively. By absorbing the kinematic recoil entirely within the $(a, r)$ dipole, this procedure ensures exact four-momentum conservation and satisfies the on-shell conditions for all resulting particles without disturbing the kinematics of previously generated, harder emissions.

The evolution scale $t$ and the light-cone momentum fraction $z$ are first mapped to the invariant mass squared $M_{bc}^2$ of the decaying system:
\begin{align}
M_{bc}^2 = \frac{t}{z(1-z)} + \frac{m_b^2}{z} + \frac{m_c^2}{1-z}~.
\end{align}
This mapping is highly non-trivial due to the finite masses involved, strictly restricting the kinematically allowed limits for $z$ and $t$ (the exact derivation of these phase space boundaries is detailed in Appendix~\ref{sec:app1}). 
In the Dipole Center-of-Mass (CM) frame, where the total invariant mass is $\sqrt{s}$, the energy $E_{bc}$ and the axial momentum $P_z$ of the $bc$-system are determined by standard two-body kinematics:
\begin{align}
E_{bc} = \frac{s + M_{bc}^2 - m_r^2}{2\sqrt{s}}, \quad P_z = \sqrt{E_{bc}^2 - M_{bc}^2}~.
\end{align}

Inside the $bc$ system of invariant mass $M_{bc}$, the two daughters are first assigned temporary momenta in the collinear direction. The initial transverse momentum magnitude and longitudinal momenta read
\begin{align}
p_\perp^{(0)} &= \sqrt{ \frac{M_{bc}^2}{P_z^2}\Bigl[E_{bc}^2 ~ z(1-z) - \tfrac14 M_{bc}^2\Bigr] }, \\
p_{z,b}^{(0)} &= \frac{E_{bc}^2 z - \tfrac12 M_{bc}^2}{P_z}, \qquad
p_{z,c}^{(0)} = \frac{E_{bc}^2 (1-z) - \tfrac12 M_{bc}^2}{P_z}.
\end{align}
These expressions follow from the requirement that the $bc$ system has invariant mass $M_{bc}$ and that the splitting variable $z$ corresponds  to the light-cone momentum fraction in the dipole CM frame.

When the daughters are massive ($m_b, m_c > 0$), the momenta obtained above do not yet satisfy the individual on-shell conditions $p_b^2 = m_b^2$ and $p_c^2 = m_c^2$. To enforce them, two rescaling factors are computed from the Kallen  function $\lambda(M_{bc}^2, m_b^2, m_c^2)$:
\begin{equation}
k_b = \frac{M_{bc}^2 - \sqrt{\lambda} + m_c^2 - m_b^2}{2M_{bc}^2}, \qquad
k_c = \frac{M_{bc}^2 - \sqrt{\lambda} + m_b^2 - m_c^2}{2M_{bc}^2}.
\end{equation}
The physical transverse momentum is obtained by scaling the temporary value,
\begin{equation}
p_\perp = p_\perp^{(0)} \times (1 - k_b - k_c),
\end{equation}
while the longitudinal momenta are shifted so that each daughter acquires the correct share of the invariant mass:
\begin{align}
\delta P_z &= k_b ~p_{z,b}^{(0)} - k_c ~p_{z,c}^{(0)}, \\
p_{z,b} &= p_{z,b}^{(0)} - \delta P_z, \qquad
p_{z,c} = p_{z,c}^{(0)} + \delta P_z.
\end{align}
This linear combination guarantees $p_b^2 = m_b^2$ and $p_c^2 = m_c^2$ while conserving the total longitudinal momentum of the $bc$ system, i.e.\ $p_{z,b} + p_{z,c} = P_z$. In the massless limit, $k_b = k_c = 0$ and the rescaling reduces to the identity.

After sampling a random azimuthal angle $\phi$, the full three-momenta of $b$, $c$, and the spectator $r$ are constructed in the dipole CM frame. The spectator receives only a longitudinal recoil, $\mathbf{p}_{r'} = -P_z  \hat{\mathbf{z}}$, reflecting the fact that in the collinear splitting picture all transverse momentum is balanced between the two daughters. The complete kinematics is then transformed back to the laboratory frame via the inverse Lorentz boost of the original dipole. The spectator $r$ is updated in place with this boosted momentum, absorbing the longitudinal recoil while the rest of the event remains unchanged.

\end{enumerate}

\section{Lund jet plane and Network architecture} \label{sec:4}

To construct the datasets for our analysis, we simulate dark dijet events initiated by the s-channel production of a dark vector boson mediator, $e^+ e^- \to Z_v \to q_D \bar{q}_D$, at a center-of-mass energy of $\sqrt{s} = 2$~TeV (with $m_{Z_v} = 2$~TeV). This hard scattering process is generated using the Pythia8 Hidden Valley module~\cite{Carloni:2010tw,Carloni:2011kk}, which serves as the physical interface to seed our parton shower. To ensure well-collimated jet topologies suitable for analysis, we impose a hard phase-space cut requiring the transverse momentum of the outgoing dark quarks to satisfy $p_T > 500$~GeV. We systematically investigate six dark gauge symmetries, including $SU(2)$, $SU(4)$, $Sp(4)$, $Sp(8)$, $SO(5)$, and $SO(7)$). Each case is simulated under four different dark gauge boson mass benchmarks: $m_V = 0$ (massless), $0.1$~GeV, $0.5$~GeV, and $1.0$~GeV. The dark quarks are assigned a small physical mass of $m_{q_D} = 10$~MeV to act as a light-flavor benchmark. For each combination of gauge group and mass benchmark, a dataset containing $5 \times 10^5$ events is generated.

Using these highly boosted initial partons, the subsequent shower evolution is simulated within our framework. The maximum evolution scale of the shower is defined by the initial hard-scattering scale, $t_{\text{max}} = s/4 = 1.0~\text{TeV}^2$. The infrared cutoff scale, $t_{\text{min}}$, acts as the non-perturbative threshold and is defined dynamically for each gauge group as the scale where the running coupling reaches the strongly coupled boundary $\alpha = 1.0$, assuming a dark confinement scale $\Lambda = 0.1$~GeV and $N_f = 5$ active flavors. The calculated $\sqrt{t_{\text{min}}}$ values for each dark gauge group are summarized in Table~\ref{tab:tmin}. To ensure numerical convergence to the continuous evolution limit, the logarithmic interval $[\ln t_{\text{min}}, \ln t_{\text{max}}]$ is discretized into $10000$ steps.

\begin{table}[htbp]
\centering
\begin{tabular}{c|cccccc}
\hline
& $SU(2)$ & $SU(4)$ & $Sp(4)$ & $Sp(8)$ & $SO(5)$ & $SO(7)$ \\
\hline
$\sqrt{t_{\text{min}}}$ (GeV) & $0.4810$ & $0.1741$ & $0.2269$ & $0.1520$ & $0.4263$ & $0.1714$ \\
\hline
\end{tabular}
\caption{Calculated infrared cutoff scales $\sqrt{t_{\text{min}}}$ for each dark gauge group, evaluated at the strongly coupled boundary $\alpha = 1.0$ under a dark confinement scale $\Lambda = 0.1$~GeV and $N_f = 5$ active flavors.}
\label{tab:tmin}
\end{table}

\subsection{The Lund Jet Plane representation}
The LJP provides a powerful theoretical and visual framework for deconstructing the substructure of jets. Grounded in the kinematics of parton branching, the LJP maps the emission history of a jet onto a two-dimensional phase space, effectively separating the perturbative radiation from non-perturbative dynamics. To construct the primary LJP, the final-state jet constituents are typically reclustered using the Cambridge/Aachen (C/A) algorithm~\cite{Dokshitzer:1997in}, which strictly enforces angular ordering. The resulting clustering tree is then iteratively declustered, following the hardest branch at each step. Each declustering node represents a $1 \to 2$ splitting and is recorded in the LJP as a coordinate pair: the logarithm of the emission angle, $\ln(1/\Delta)$, and the logarithm of the relative transverse momentum of the emission, $\ln(k_T)$.

In this representation, the soft and collinear singularities governed by the Altarelli-Parisi splitting functions explicitly manifest as a uniform probability density in the perturbative regime~\cite{Lifson:2020gua}. 
By organizing emissions in this manner, the LJP isolates critical dynamical features of the shower, including the running of the strong coupling, Casimir scaling, and Sudakov suppression, while simultaneously confining non-perturbative hadronization effects to the low-$k_T$ lower boundary of the plane~\cite{Dreyer:2018nbf,Cohen:2023mya}. 
Owing to its ability to encode the complete branching history and isolate physical regimes, the LJP has emerged as a highly sensitive basis for probing jet dynamics and has been successfully integrated with advanced machine learning architectures, such as Graph Neural Networks (GNNs), for complex jet tagging tasks~\cite{Dreyer:2020brq}.

\begin{figure}[htbp]
\includegraphics[width=0.48\textwidth]{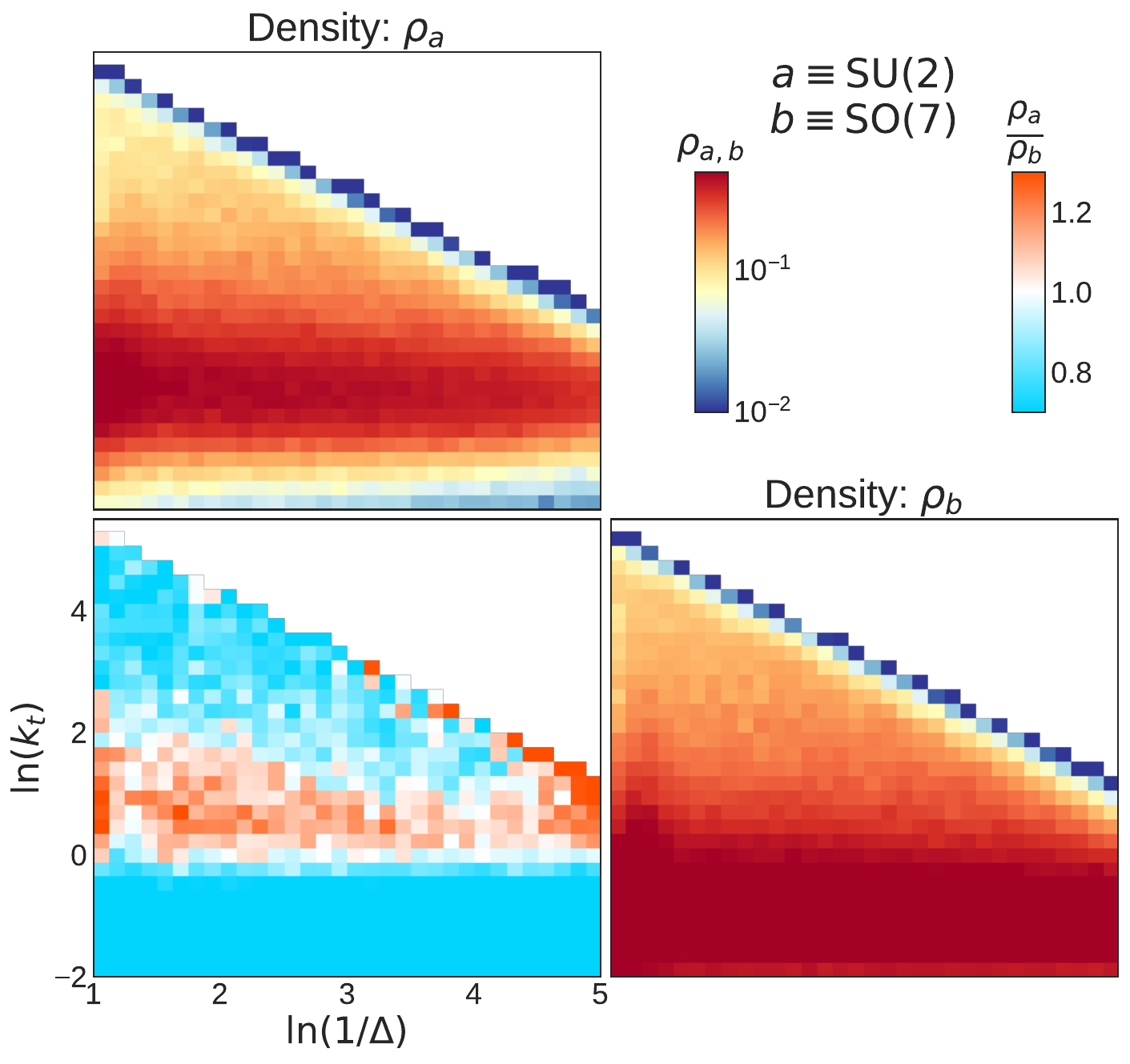}
\includegraphics[width=0.48\textwidth]{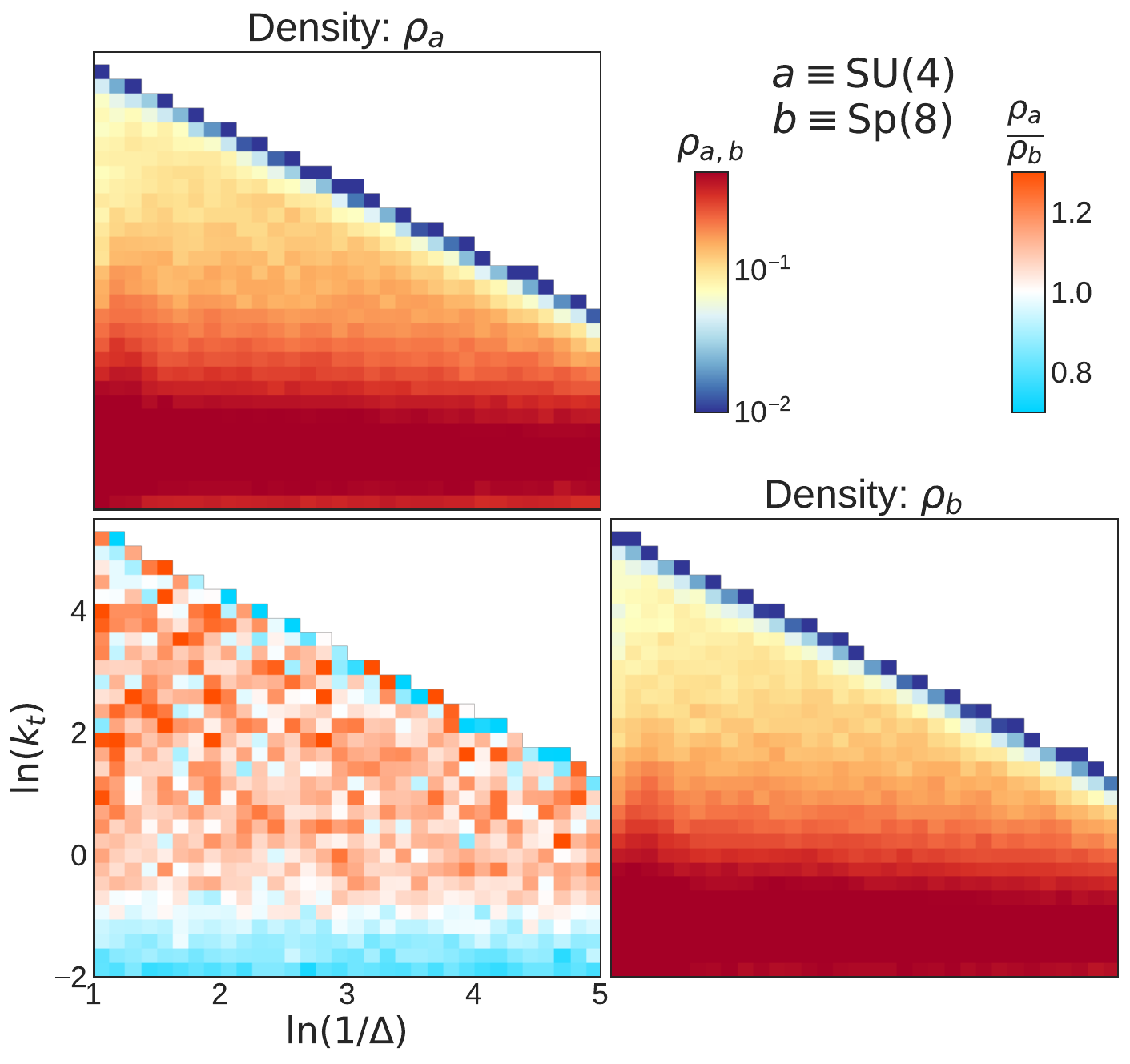}
\caption{Primary LJP density distributions and their ratios for jets generated with massless dark gauge bosons under different gauge symmetries. The left panel compares the $SU(2)$ and $SO(7)$ group, illustrating a distinct scale-dependent emission ratio. The right panel compares the $SU(4)$ and $Sp(8)$ group, showing more uniform and stable radiation ratio in high $k_T$ regime. 
\label{fig:ljp}}
\end{figure}

To investigate the dynamical differences between dark sector gauge symmetries, we analyze the Primary LJP densities for jets generated with massless gauge boson in 
$SU(2)$, $SU(4)$, $SO(7)$ and $Sp(8)$ theories as illustrative benchmarks. As shown in Figures~\ref{fig:ljp}, the emission patterns exhibit a hierarchical structure that extends beyond simple color-charge scaling. 
To leading order, the primary intensity of the radiation density in the LJP is given by $\rho \simeq \frac{2\alpha(k_T)}{\pi} C_F$. While the overall magnitude is governed by the fundamental Casimir factor $C_F$, the detailed LJP distributions reveal deeper insights into the parton shower evolution and the explicit running behavior of the dark coupling $\alpha(k_T)$.

In the comparison between $SU(2)$ and $SO(7)$, the ratio plot reveals a distinct scale-dependent behavior. 
Radiation in $SU(2)$ is significantly suppressed at both the high and very low $k_T$ regions due to its much smaller color charge ($C_F = 3/4$ for $SU(2)$ versus $C_F = 3$ for $SO(7)$). However, it exhibits a prominent enhancement band relative to $SO(7)$ at intermediate scales, specifically localized around $\ln(k_T) \sim 0\text{--}2$. Because the ratio of the Casimir invariants $C_{F, SU(2)} / C_{F, SO(7)}$ is a constant, this non-uniform feature directly maps the relative differences in the $\beta$-functions of the two groups. It suggests that the steeper running of the $SU(2)$ dark coupling $\alpha(k_T)$ at these specific energy scales effectively compensates for its lower color charge, momentarily overtaking the emission intensity of the $SO(7)$ shower.
In contrast, the comparison between $SU(4)$ and $Sp(8)$ shows a much more uniform and stable ratio across the entire perturbative regime. For these two groups, both the fundamental Casimir invariants ($C_F = 1.875$ for $SU(4)$ and $C_F = 2.25$ for $Sp(8)$) and their $\beta$-function coefficients are structurally similar. This relative lack of contrast suggests a convergence in radiation patterns for higher-rank gauge groups, where the differences in group-theoretic factors become less dominant in shaping the overall jet substructure.

Furthermore, the non-uniform color gradients observed in the ratio plots provide distinct evidence of differential Sudakov resummation effects. In the LJP framework, each emission point represents a branching whose probability is modulated by the Sudakov form factor, which is defined as the probability that the parton evolved from the hard scale down to the emission scale without radiating. Gauge groups with larger color factors or stronger initial couplings radiate more profusely at early, large-angle scales, leading to a faster depletion of the Sudakov survival probability at later stages. The observed gradients indicate that the evolution of the initial parton accumulates these effects at different rates across gauge symmetries, shifting the bulk of the radiation density to different regions of the phase space. 

Finally, the transition from the perturbative, uniformly populated `triangle' down to the non-perturbative regime, as previously discussed in Ref.~\cite{Cohen:2023mya}, is clearly visible across all models. Together, these features demonstrate that the LJP, together with the ratio of LJP densities, serves as a highly sensitive physical probe for discriminating dark sector symmetries through their explicit dynamical shower footprints.

\subsection{Network Architecture}

While the Lund Jet Plane distributions provide clear visual evidence of gauge-dependent radiation patterns, systematically extracting these intertwined, multi-dimensional signatures for event-by-event classification requires sophisticated multivariate analyses. 
To achieve robust discrimination between the underlying dark gauge symmetries, we have developed a specialized deep learning architecture that directly processes the branching history of the shower. The data preparation pipeline translates the raw final-state kinematics into this structured representation. 

Specifically, the jet deconstruction process begins by clustering final-state particles using the anti-$k_t$ algorithm ($R=0.4$). To obtain a proxy for the QCD branching history, jet constituents are reclustered using the C/A algorithm. Each internal node of the resulting C/A tree is declustered into two branches, $j_1$ and $j_2$ ($p_{T,1} > p_{T,2}$), for which five primary Lund coordinates are extracted: $(\ln k_t, \ln (1/\Delta), \ln z, \ln m, \psi)$. 
These declustering variables are defined as follows:
\begin{align}
&\Delta\equiv\sqrt{(y_1 - y_2)^2 + (\phi_1-\phi_2)^2}, \quad k_t \equiv p_{T,2} \Delta, \quad z\equiv\frac{p_{T,2}}{p_{T,1}+p_{T,2}}, \\
&m^2\equiv(p_1^\mu +p_2^\mu)^2, \quad \psi\equiv \tan^{-1}\frac{y_2-y_1}{\phi_2-\phi_1},
\end{align}
where $y_i$, $\phi_i$, $p_{T,i}$ and $p^\mu_i$ denote rapidity, azimuthal angle, transverse momentum, and four momentum of the $i$-th subjet, respectively. 
To maintain full kinematic fidelity, the 4-momentum $(p_x, p_y, p_z, E)$ of the emitted branch ($j_2$) is recorded for each splitting. The azimuthal angle $\psi$ is transformed into $(\cos \psi, \sin \psi)$ to account for its periodicity. This yields a comprehensive $10$-dimensional feature vector for each of the $L$ nodes in the jet tree.

\begin{figure}[htbp]
\centering
\includegraphics[width=0.8\textwidth]{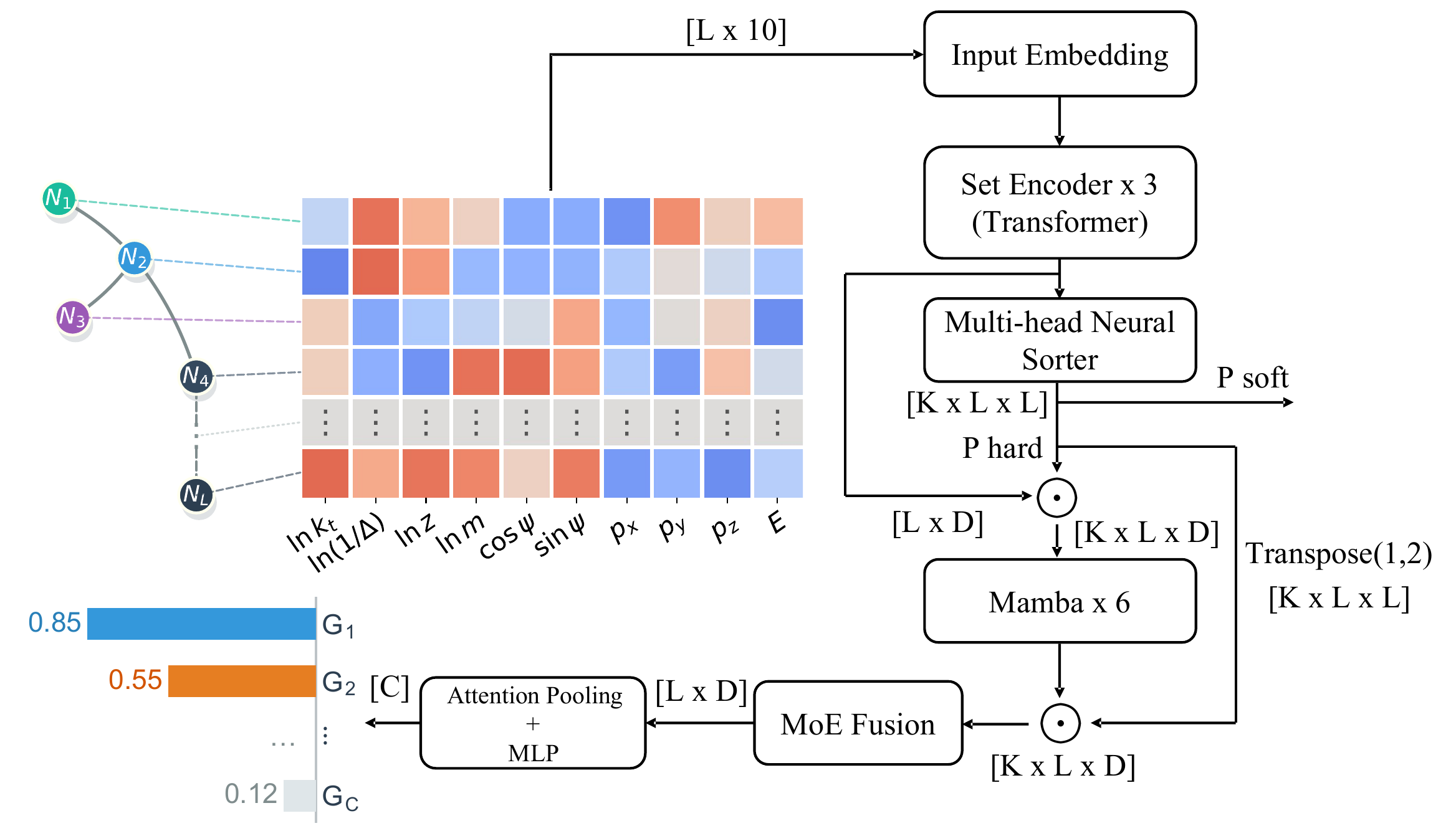}
\caption{Schematic overview of the NS-MambaNet architecture for jet classification.}
\label{fig:model}
\end{figure}

The deep learning architecture, illustrated in Figure~\ref{fig:model}, is specifically designed to respect the underlying symmetries of the physical data while capturing complex, sequential shower correlations. The jet nodes are initially represented as an unordered set of shape $[L \times 10]$, reflecting the permutation invariance of the radiation pattern. An Input Embedding layer projects each node's $10$-dimensional feature vector into a $D$-dimensional latent space, yielding a tensor of shape $[L \times D]$. This representation is then processed by a Set Encoder, comprising 3 layers of Transformer-based self-attention blocks without positional encodings. This permutation-equivariant module models the global, long-range correlations among all emissions within the jet, ensuring that the jet-level representation facilitates information exchange across all nodes without imposing an artificial sequence.

Because parton showers are fundamentally sequential cascade processes, the model must transition from an unordered set to sequential representations. This is achieved via a Multi-Head Neural Sorter. This module learns $K$ optimal sorting strategies by predicting scalar scores for each node. An iterative Sinkhorn operator converts these scores into a soft, differentiable permutation matrix $P_{\text{soft}}$ of shape $[K \times L \times L]$. Multiplying the latent set $[L \times D]$ by $P_{\text{soft}}$ transforms the data into $K$ independent, dynamically ordered sequences of shape $[K \times L \times D]$. 

These $K$ sequences are subsequently processed by a shared sequence-modeling backbone consisting of 6 Mamba layers. By leveraging selective State Space Models, the Mamba backbone efficiently captures structural, causal correlations within the sorted emission trees with linear time complexity, which is a significant advantage over standard Transformers for long fragmentation cascades. To restore the permutation-invariant context of the original set, the sequence features are mapped back to their original unordered indices. As shown in the figure, this is accomplished by taking the transpose of the hard permutation matrix $P_{\text{hard}}$ (producing a $[K \times L \times L]$ inverse mapping) and multiplying it with the Mamba output, yielding an unsorted feature tensor of shape $[K \times L \times D]$.

Finally, the representations from the $K$ distinct sorting heads are integrated via a Gated Fusion module based on a Mixture-of-Experts (MoE) mechanism. This module dynamically computes attention weights to consolidate the node-level features across the $K$ heads, collapsing the tensor back to a unified $[L \times D]$ representation. The global jet-level embedding is then obtained through Query Attention Pooling, where a learnable query vector aggregates the $L$ nodes based on their learned relevance for class discrimination. The resulting pooled vector is passed through a Multi-Layer Perceptron (MLP), which projects it to the final $[C]$-dimensional output space representing the distinct gauge symmetry classes. The entire architecture is optimized end-to-end using a Binary Cross Entropy loss function.

\section{Decoding the gauge structure} \label{sec:5}

\begin{figure}[htbp]
\centering
\includegraphics[width=0.45\textwidth]{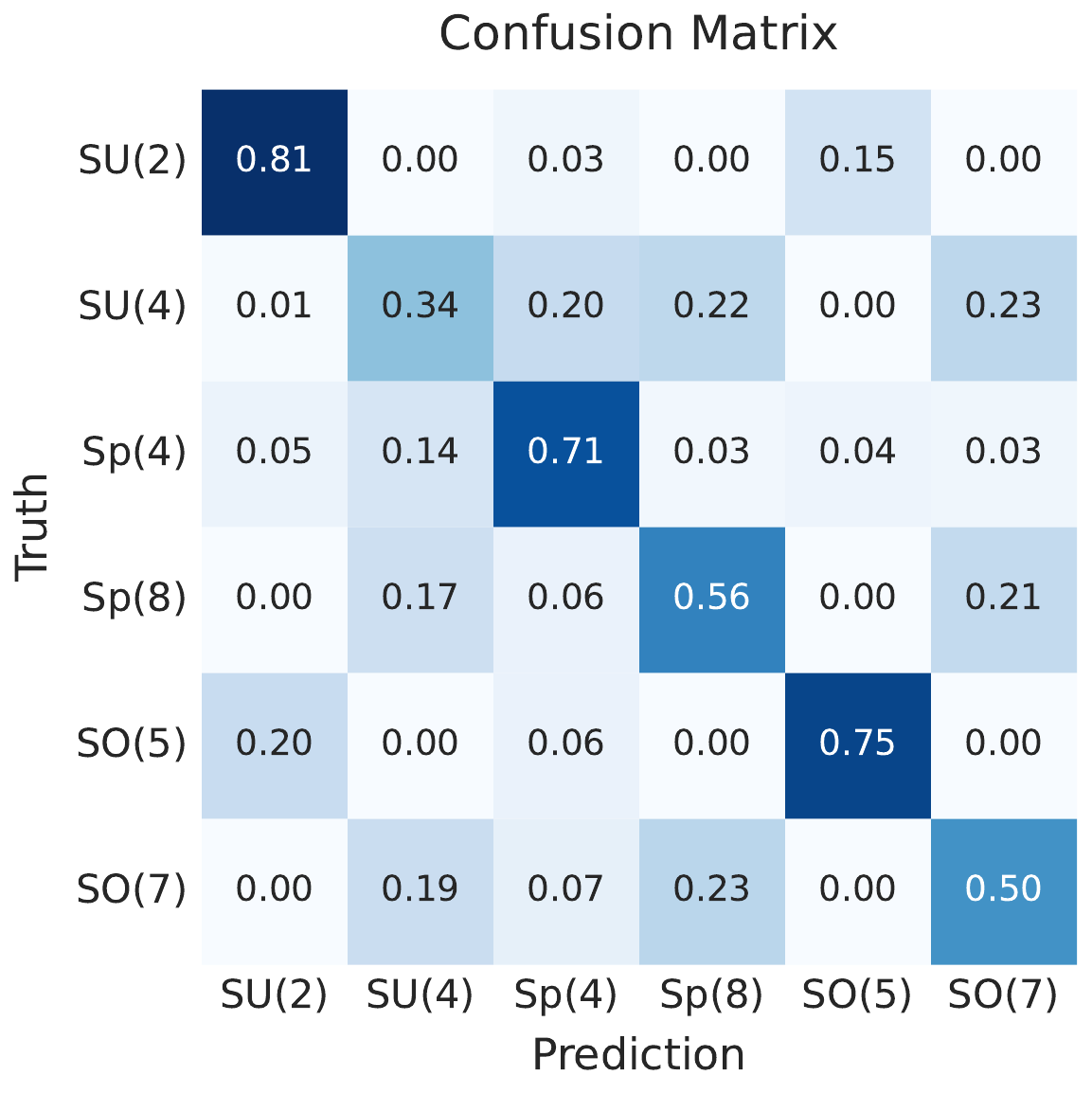}
\includegraphics[width=0.45\textwidth]{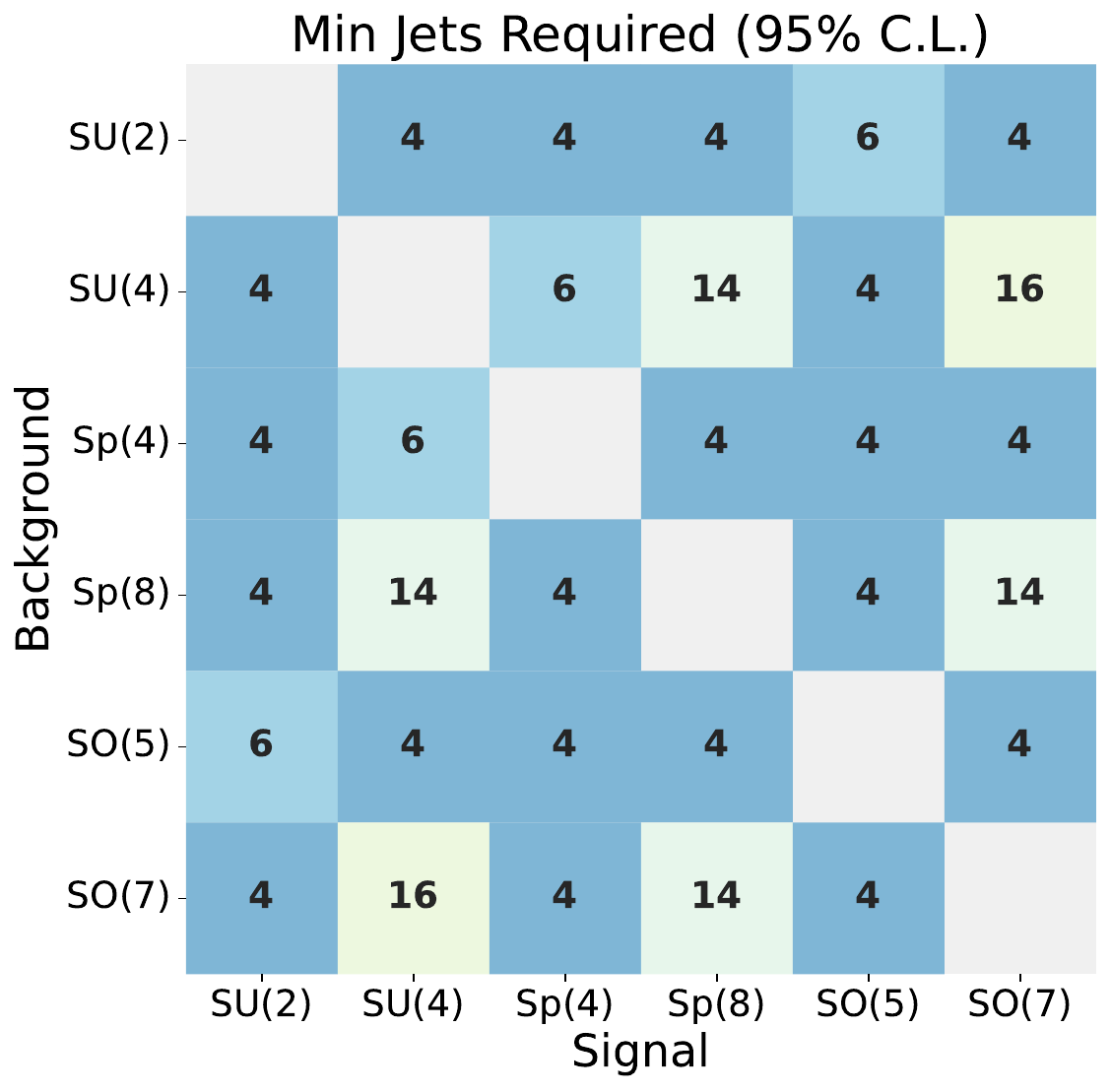}
\caption{Baseline classification performance of the NS-MambaNet on jets generated with massless dark gauge bosons 
The left panel shows the multi-class confusion matrix evaluated on a single-jet basis.
The right panel displays the minimum number of jets required to achieve a 95\% C.L. discrimination between any pair of gauge symmetries. }
\label{fig:cm}
\end{figure}

Before investigating mass and non-perturbative effects, we evaluate the baseline performance of the NS-MambaNet using a dataset where the dark gauge bosons are massless ($m_V = 0$). The left panel of figure~\ref{fig:cm} illustrates the multi-class confusion matrix on a single-jet basis. The network exhibits high classification accuracy for several gauge groups, notably $SU(2)$ (81\%), $Sp(4)$ (71\%), and $SO(5)$ (75\%). These groups possess distinct color factor configurations that lead to unique radiation topologies, allowing the model to identify them with high fidelity even from a single jet.
However, a degree of degeneracy is observed among $SU(4)$, $Sp(8)$, and $SO(7)$, with $SU(4)$ showing the lowest single-jet accuracy at 34\%. This confusion arises primarily from the similarities in their perturbative radiation patterns in the massless limit. Specifically, $SU(4)$ is frequently misclassified as $Sp(8)$ (22\%) or $SO(7)$ (23\%), suggesting that the single-jet substructure features learned by the model for these groups are partially overlapping.

Despite the challenges in single-jet classification, the discrimination power increases dramatically when considering the statistical accumulation of multiple jet events.
To quantify the statistical significance of hypothesis separation over a pseudo-dataset of multiple jets, we perform a frequentist hypothesis test using the log-likelihood ratio (LLR) as the test statistic. For a single jet event $x$, let $s_i(x)$ denote the neural network logit predicted for gauge symmetry hypothesis $H_i$ ($i \in \{A, B\}$). Under the assumption of equal prior probabilities, the single-event LLR is evaluated using the ratio of posterior probabilities obtained via sigmoid activations:

\begin{equation}
\mathcal{L}(x) = \ln \frac{P(x | H_A)}{P(x | H_B)} = \ln \frac{\sigma(s_A(x))}{\sigma(s_B(x))} = \ln\left(1 + e^{-s_B(x)}\right) - \ln\left(1 + e^{-s_A(x)}\right)~.~
\end{equation}

For a sample of size $N$, the cumulative test statistic $TS$ is defined as the sum of LLRs over the observed events:

\begin{equation}
TS = \sum_{i=1}^{K} \mathcal{L}(x_i)~,~
\end{equation}
where the actual number of observed jets $K$ in a given pseudo-experiment fluctuates according to a Poisson distribution, $K \sim \text{Poisson}(N \cdot \theta)$. Here, $\theta \sim \text{Lognormal}(0, \sigma_{\text{sys}})$ is a nuisance parameter introduced to model a conservative $\sigma_{\text{sys}} = 10\%$ normalization systematic uncertainty. 

To determine the minimum number of jets required for discrimination, we generate $10^5$ pseudo-experiments (toy Monte Carlo) to construct the empirical probability distributions of $TS$ under both hypotheses, $P(TS | H_A)$ and $P(TS | H_B)$. A symmetric hypothesis test is then conducted by scanning for a decision boundary $TS_{\text{cut}}$ that simultaneously constrains the Type-I error rate $E_\alpha = P(TS < TS_{\text{cut}} | H_A)$ and the Type-II error rate $E_\beta = P(TS \ge TS_{\text{cut}} | H_B)$ to be below $5\%$. The minimum expected event count $N$ that satisfies this separability condition is efficiently solved via binary search.

The right panel of figure~\ref{fig:cm} displays the minimum number of jets required to achieve a 95\% C.L. discrimination between any two gauge symmetries. We observe that even for the most "degenerate" pairs identified in the confusion matrix, the statistical significance builds up rapidly. For example, the $SU(4)$ and $Sp(8)$ pair, which exhibits severe confusion on a single-jet basis, can be cleanly distinguished with only 14 jets. The most difficult pair to separate, $SU(4)$ and $SO(7)$, requires merely 16 jets to reach the 95\% C.L. threshold.

For the majority of other pairs, such as $SU(2)$ vs.~$SO(7)$ or $Sp(4)$ vs.~$SO(5)$, the required jet count remains as low as 4. This demonstrates that the NS-MambaNet is capable of extracting subtle, gauge-dependent signatures. 
While these signatures may be masked by physical fluctuations in a single event, they become statistically distinguishable even in small data samples. 
This baseline performance confirms the high efficiency of the neural sorting and Mamba2-based sequential modeling architecture in characterizing the dark sector gauge structure from final-state jet observables.

\subsection{Mass effects in the dark shower}

\begin{figure}[htbp]
\includegraphics[width=0.48\textwidth]
{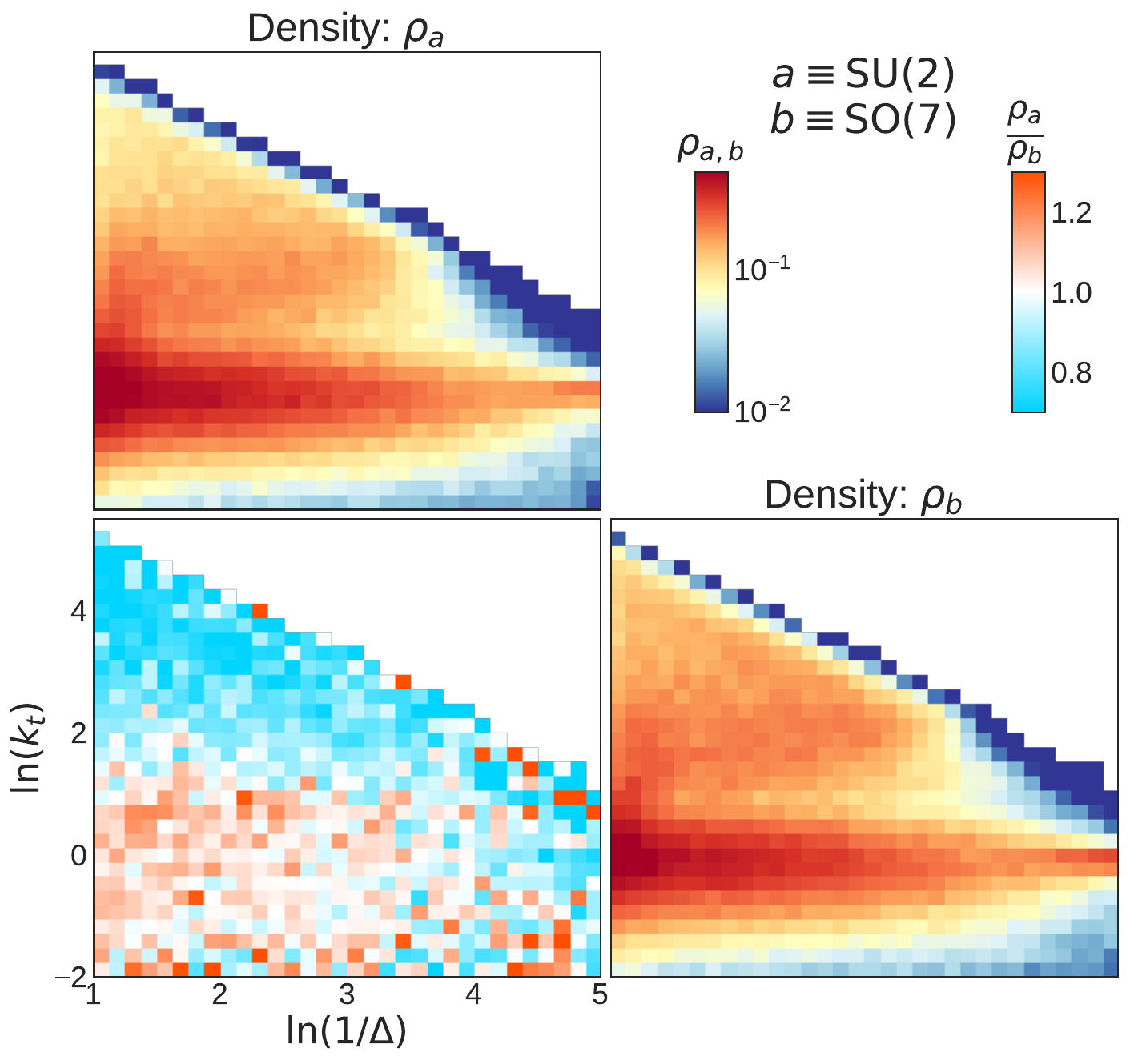}
\includegraphics[width=0.48\textwidth]
{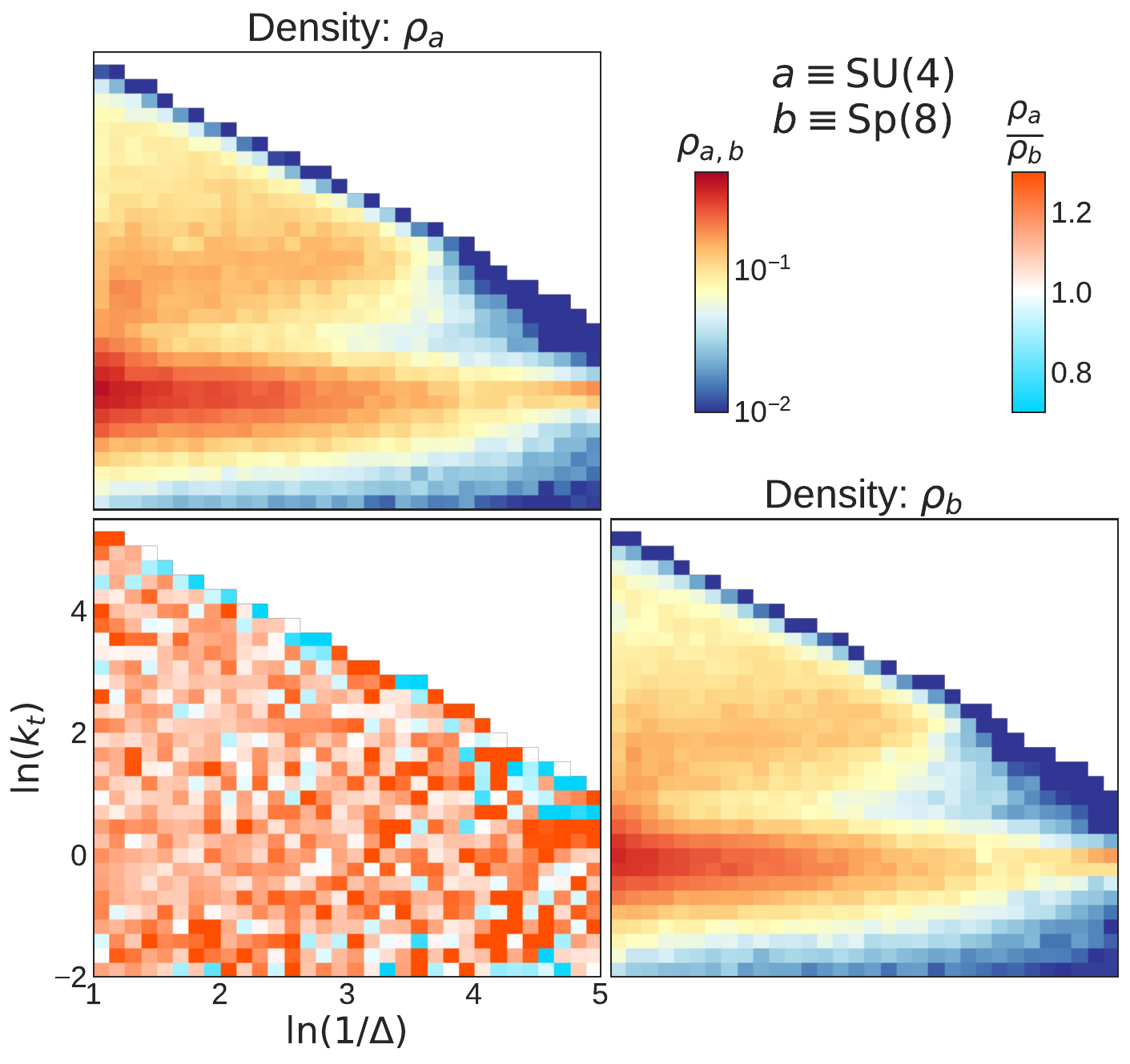}
\caption{Primary LJP density distributions and their ratios for jets generated with massive dark gauge bosons ($m_V=1$ GeV). 
The same representative gauge groups as in Figure~\ref{fig:ljp} are chosen for comparison. 
\label{fig:mass_ljp}}
\end{figure}

The inclusion of a finite dark gauge boson mass $m_V = 1$~GeV significantly alters the topology of the Lund jet plane, as shown in the density distributions of Figure~\ref{fig:mass_ljp}. A striking feature is the emergence of a clear kinematic bifurcation near $\ln(k_T) \approx 0$ (corresponding to the mass scale $m_V$). Below this threshold ($\ln(k_T) \lesssim 0$), a high-density radiation region appears that is visually decoupled from the perturbative shower triangle at higher scales ($\ln(k_T) \gtrsim 0$). This separation indicates that the emission of massive mediators leads to a distinct "soft radiation island", where the phase space is no longer governed by the standard collinear-soft singularities but is instead dominated by the massive splitting kernels.

The impact of this mass scale is even more pronounced in the LJP ratio plots. Compared to the massless case, where the emission ratio is typically suppressed for smaller groups across the infrared (IR) regime, the massive results show a significant increase in the ratio within the deep IR region ($\ln(k_T) < -1$), indicated by the prevalent orange-red hues. This suggests that the dark boson mass acts as an infrared regulator that differently affects the radiation of each gauge symmetry. Specifically, the relative enhancement in the ratio at very low $k_T$ highlights that as the shower enters the massive radiation regime, the differences in Casimir scaling are partially overshadowed by the mass-induced threshold effects. Consequently, this provides a unique experimental signature that could be used to calibrate the mass spectrum of the dark sector.

\begin{figure}[htbp]
\includegraphics[width=0.32\textwidth]{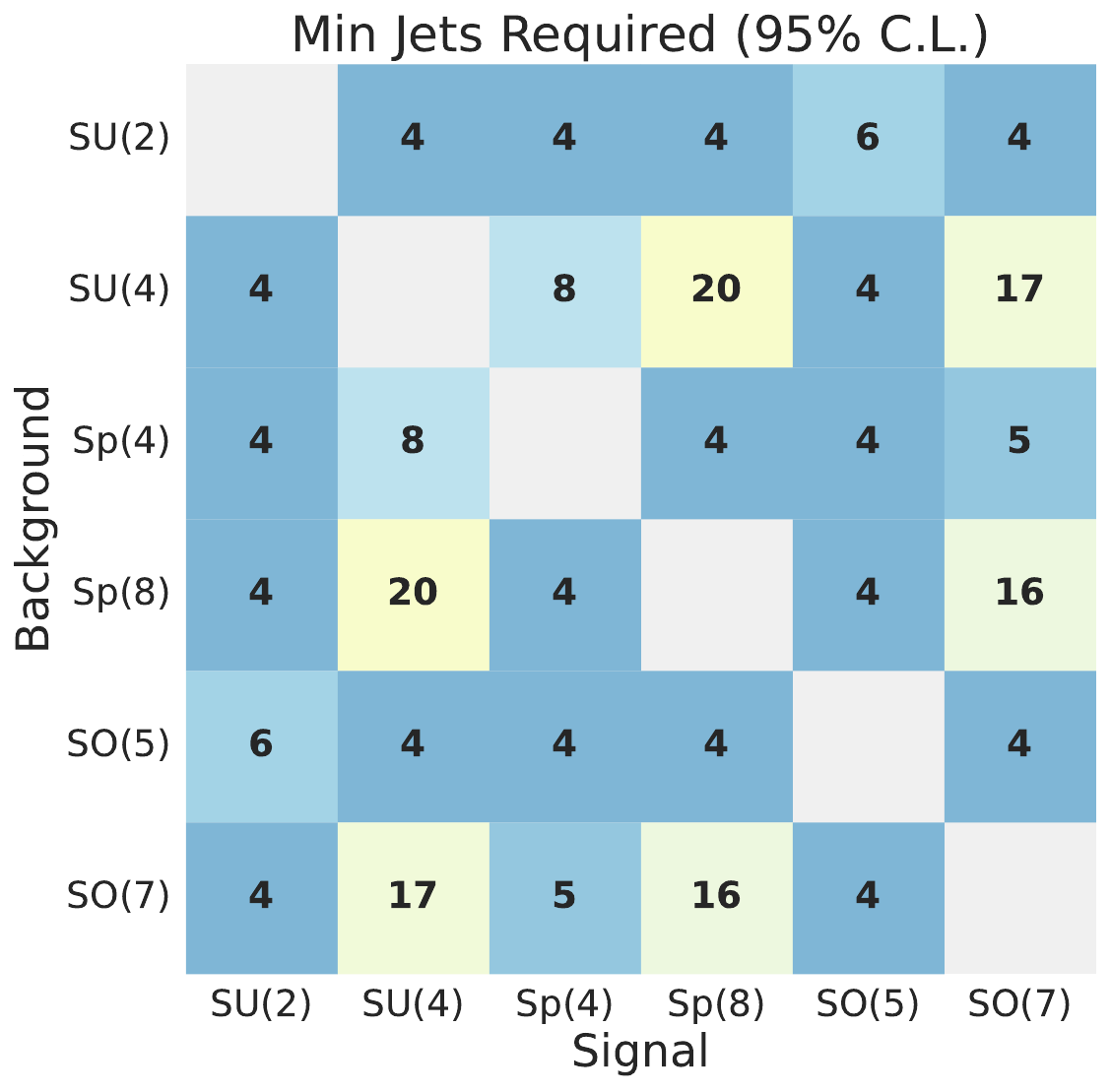}
\includegraphics[width=0.32\textwidth]{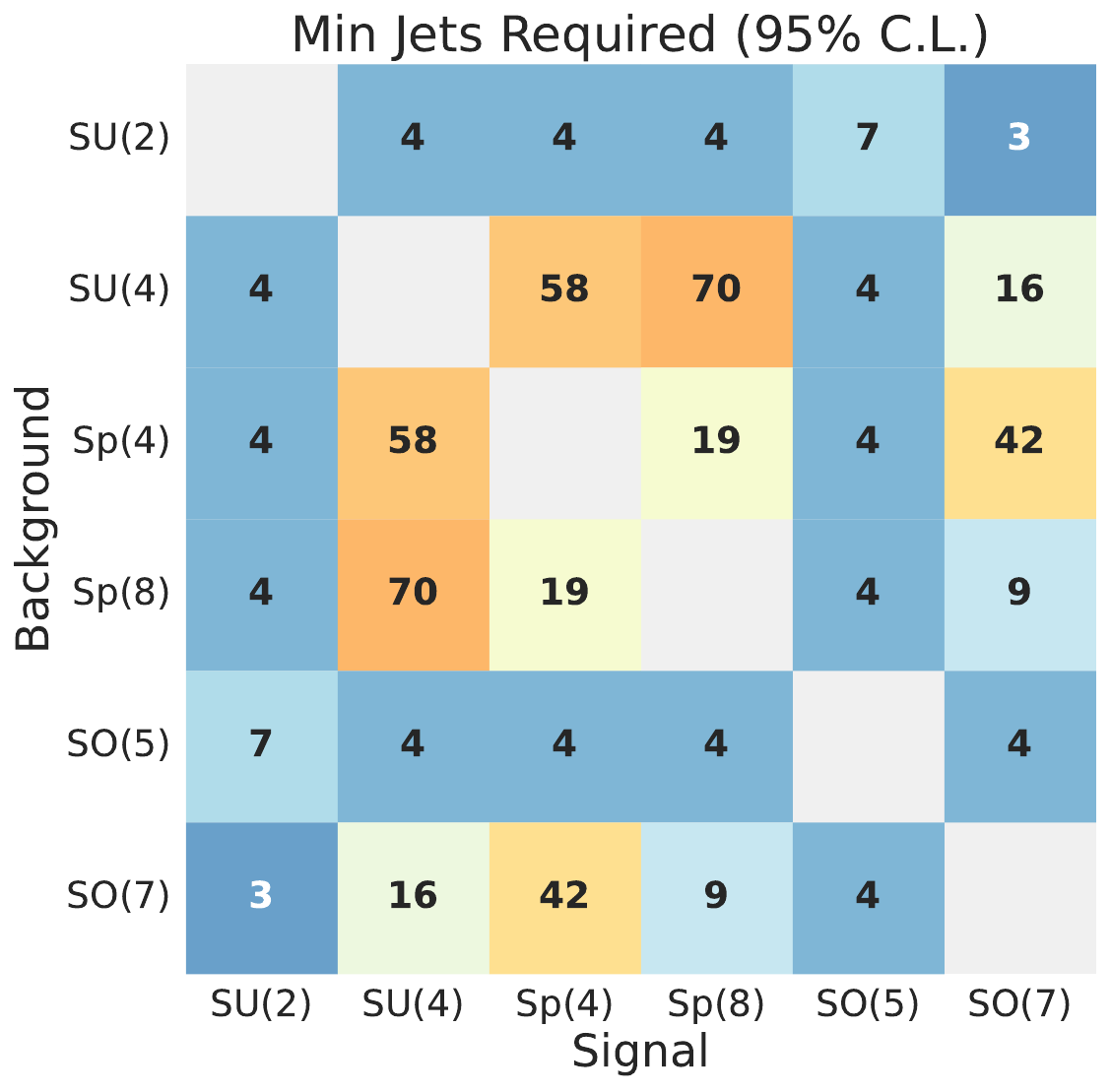}
\includegraphics[width=0.32\textwidth]{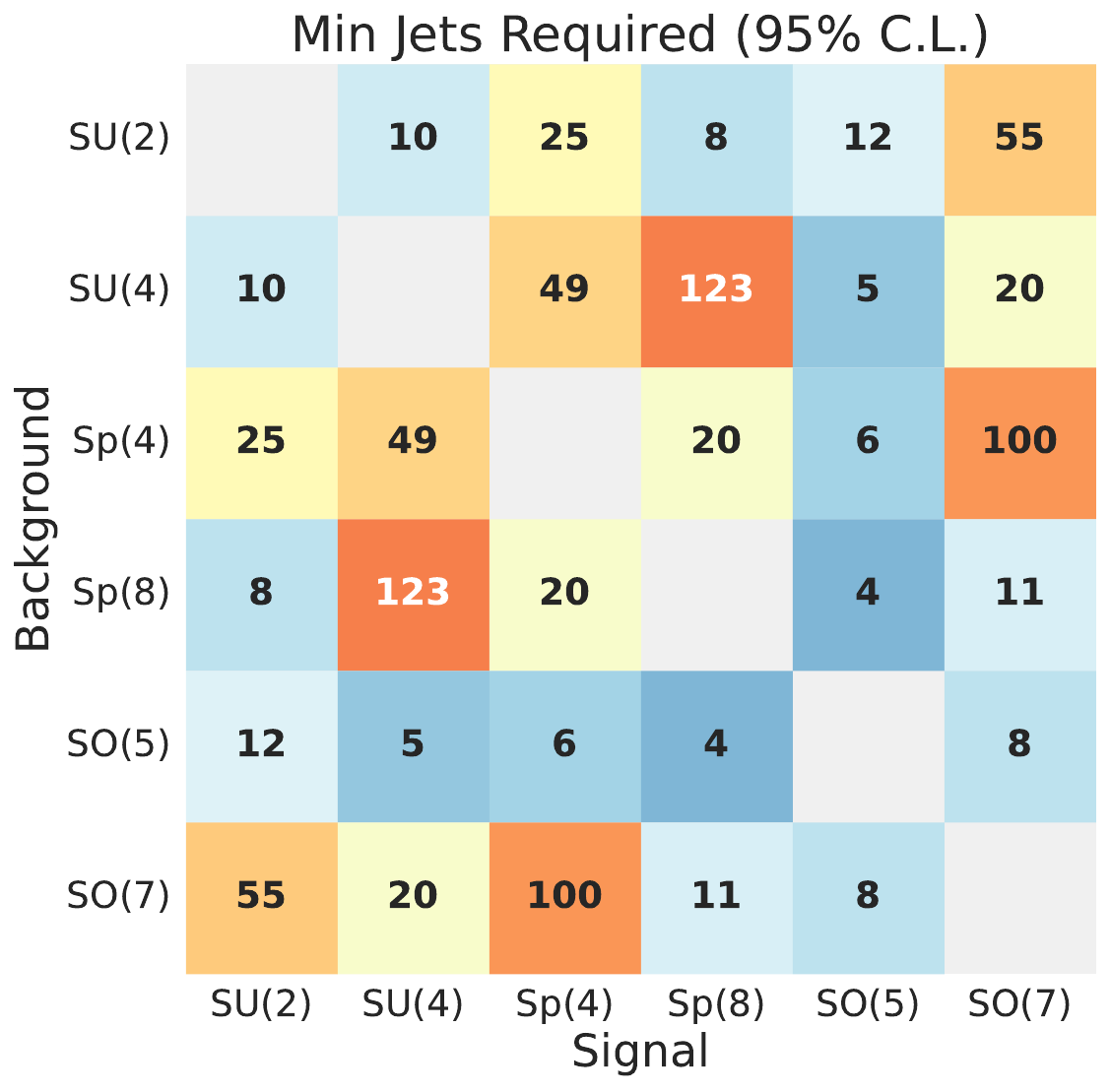}
\caption{Minimum number of jets required to achieve a 95\% C.L. discrimination between different pairs of dark sector gauge symmetries. The panels from left to right correspond to gauge boson masses of 0.1 GeV, 0.5 GeV, and 1.0 GeV, respectively.
\label{fig:mass_njet}}
\end{figure}

To evaluate how the mass scale of the dark gauge bosons affects classification, Figure~\ref{fig:mass_njet} illustrates the minimum number of jets required for a 95\% C.L. separation at $m_V = 0.1$~GeV, $0.5$~GeV, and $1.0$~GeV. We observe a systematic upward trend in the required jet counts as the gauge boson mass increases. While most symmetry pairs are easily separable with fewer than 10 jets in the lightest scenario ($m_V = 0.1$~GeV), heavier masses significantly degrade this performance. For instance, the required jet count for the challenging $SU(4)$ vs. $Sp(8)$ pair rises from 20 to 123 as the mass increases from 0.1~GeV to 1.0~GeV, while a previously easy pair like $SU(2)$ vs. $SO(7)$ increases from 4 to 55 jets.

This degradation in discrimination power at higher mass scales is primarily driven by two facts. First, as analyzed in Appendix~\ref{sec:app1}, increasing the gauge boson mass $m$ dynamically shrinks the kinematically allowed phase space. In this mass-dominated regime, the purely group-theoretic differences between the Casimir invariants ($C_A$ vs.~$C_F$) are increasingly masked by the dominant kinematics, meaning their distinct scaling behaviors are no longer the primary driver for discrimination. Second, this phase-space suppression reduces the average branching multiplicity inside the jet. With fewer emissions occurring, the total amount of statistical information carried by each jet decreases, consequently requiring significantly larger event samples to achieve the same confidence level.

\subsection{Robustness Against Non-Perturbative Effects}

\begin{figure}[thbp] \centering
\includegraphics[width=0.45\textwidth]{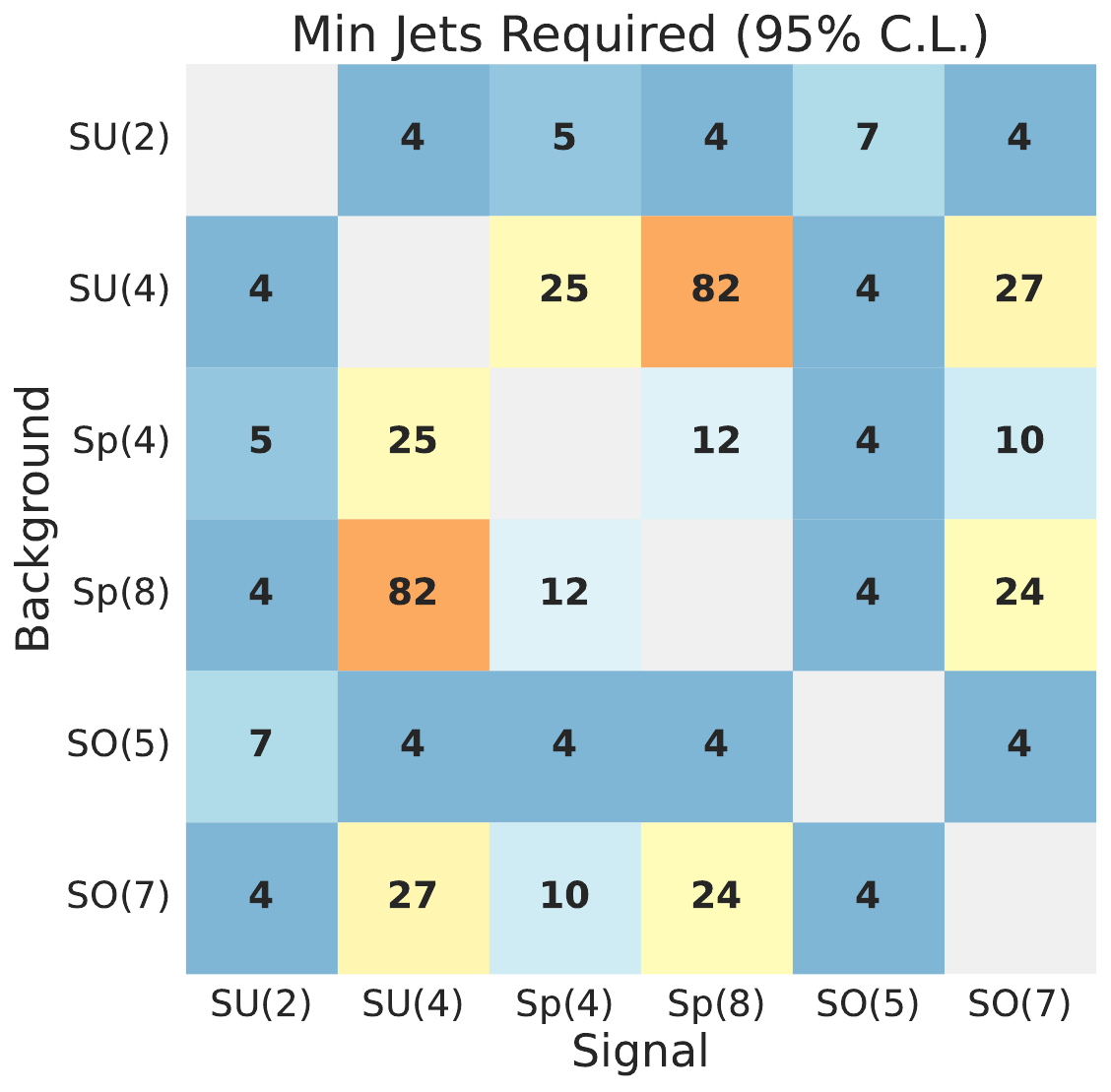}
\includegraphics[width=0.45\textwidth]{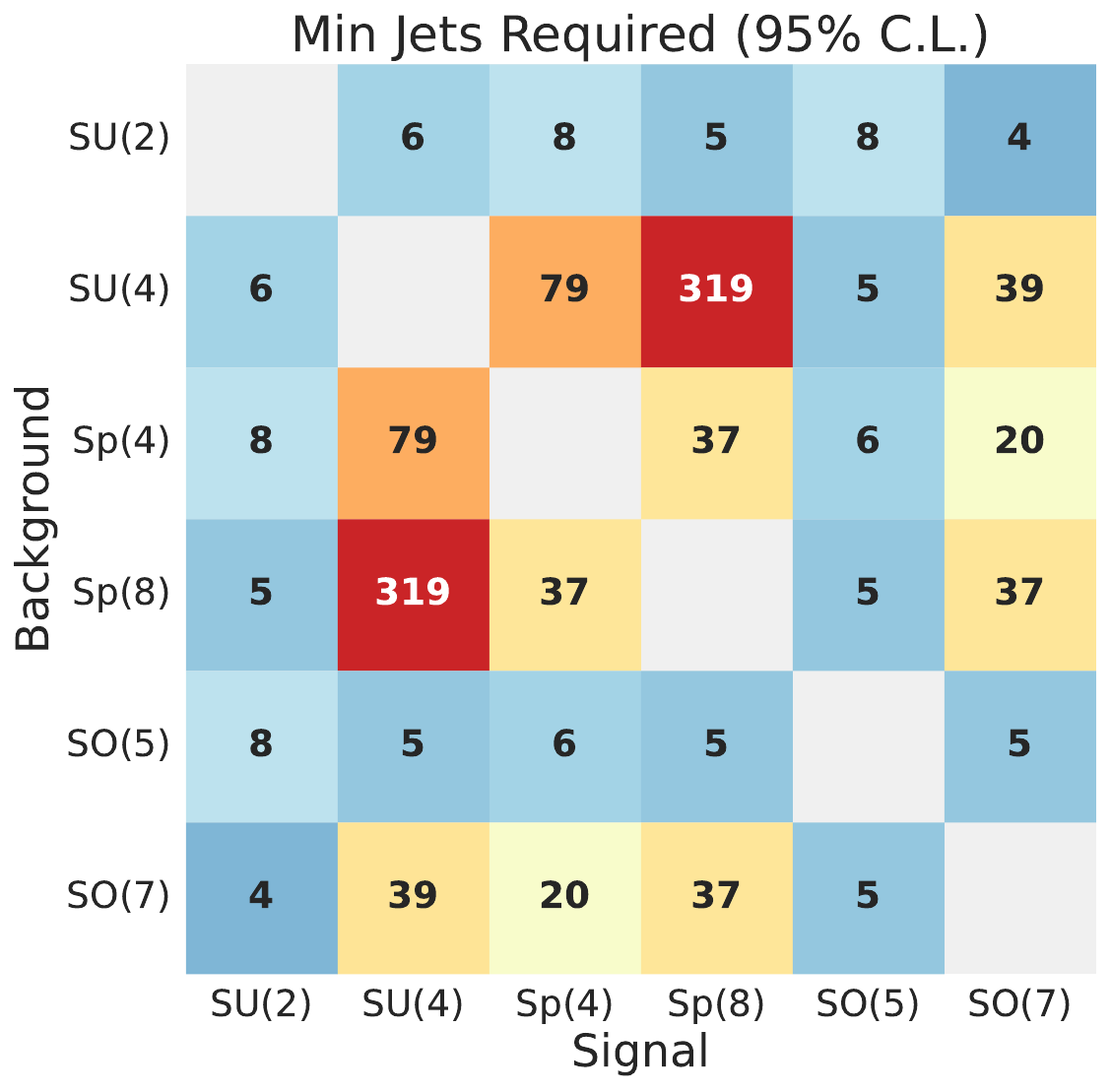}
\caption{The minimum number of jets required to achieve a 95\% C.L. separation between gauge symmetries when applying a transverse momentum cutoff of $ k_T>0.5$ GeV (left panel) and $k_T>1.0$ GeV (right panel). The dark gauge boson is taken to be massless. 
\label{fig:kt_njet}}
\end{figure}

To evaluate the resilience of the learned representations against the soft and non-perturbative effects, and to assess the model's performance across different physical regimes, the same analysis framework is applied to jet samples using various infrared cutoffs on the emission transverse momentum, $k_T$. 
We investigate three distinct kinematic selection scenarios: an inclusive selection ($k_T > 0$) that incorporates the full range of radiation, including the low-$k_T$ region dominated by non-perturbative effects, and two hard-scale cutoffs ($k_T > 0.5$~GeV and $k_T > 1.0$~GeV, as shown in Figure~\ref{fig:kt_njet}) designed to systematically filter out soft emissions. By progressively raising this threshold, we quantify the extent to which the network relies on infrared physics, as opposed to its ability to discriminate gauge symmetries based exclusively on the harder, perturbative branching structures.

As illustrated in Figure~\ref{fig:kt_njet}, increasing the $k_T$ cutoff leads to an overall upward trend in the required number of jets, reflecting the loss of distinguishing information stored in soft radiation patterns. Nevertheless, the classifier demonstrates remarkable robustness for most symmetry pairs. For instance, discriminating $SU(4)$ from $SO(5)$, or $SU(2)$ from $SO(7)$, requires fewer than 10 jets even under the most stringent cut ($k_T > 1.0$~GeV). This robust performance stems from the distinct Casimir invariants of these groups, which govern the Altarelli-Parisi splitting kernels in the perturbative regime, {\textit{e.g.}}, $SU(4)$ has $C_F = 1.875$ and $C_A = 4$, whereas $SO(5)$ possesses $C_F = 2$ and $C_A = 3$. 
Because the relative rate of secondary gluon branchings to primary quark emissions is characterized by the ratio $C_A/C_F$ ($2.13$ for $SU(4)$ vs.~$1.50$ for $SO(5)$), these differing ratios yield distinct intra-jet multiplicities that the network efficiently captures, regardless of the soft cutoffs.

The most striking impact of the IR cutoff is observed in the discrimination of specific pairs, particularly $SU(4)$ and $Sp(8)$, which exhibit a severe degradation in performance. In the left panel of Figure~\ref{fig:kt_njet} ($k_T > 0.5$~GeV), the required number of jets to distinguish this pair is 82. As the cutoff is raised to 1.0~GeV (right panel), this number spikes sharply to 319. 
From a theoretical perspective, this difficulty in the high $k_T$ regime can be attributed to a Casimir degeneracy. In the hard-scattering limit ($k_T > 1.0$~GeV), the jet substructure is primarily dictated by the leading-order splitting probabilities, which scale with $C_F$ for $q \to qg$ and $C_A$ for $g \to gg$. For $SU(4)$, the color factors are $C_F = 15/8 = 1.875$ and $C_A = 4$, whereas for $Sp(8)$, the corresponding factors are $C_F = 9/4 = 2.25$ and $C_A = 5$. While their absolute intensities differ by approximately $20\% \sim 25\%$, the radiation pattern ratio within the jet, governed by $C_A/C_F$, is largely similar, {\textit{i.e.}} $C_A/C_F \approx 2.133$ for $SU(4)$ and $C_A/C_F \approx 2.222$ for $Sp(8)$ (a relative difference of only $\sim 4\%$). 
This implies that the topological shapes of their perturbative branching trees driven by the relative multiplicity of secondary vs.~primary emissions, are nearly indistinguishable. 
Meanwhile, differences in the absolute scale of radiation are easily washed out by statistical fluctuations or compensated by the running coupling $\alpha$.
Consequently, when the $k_T > 1.0$~GeV cut removes soft emissions, the network is forced to rely on this nearly degenerate hard "skeleton", leading to the spike in the required number of jets. Conversely, in the lower-$k_T$ and inclusive scenarios, the larger number of soft emissions provides higher statistics and exhibits stronger dependence on the absolute color factors, thereby breaking the degeneracy. Removing this soft radiation "tail" thus significantly degrades the statistical significance per jet for the $SU(4)$ vs. $Sp(8)$ pair. 

In summary, the variation in discrimination power across different cutoffs confirms that the learned features are deeply rooted in the underlying gauge-invariant branching history. While the soft emissions provide critical auxiliary information to break leading-order Casimir degeneracies (as seen in the $SU(4)/Sp(8)$ case), the model's ability to maintain high efficiency for the majority of group pairs under strict cutoffs proves its exceptional robustness in extracting fundamental perturbative signatures.

\section{Conclusion} \label{sec:6}

In this paper, we explored the perturbative radiation patterns of dark showers to discriminate the underlying gauge symmetries of a secluded dark sector. 
To achieve this, we developed a generalized Monte Carlo parton shower algorithm capable of accommodating arbitrary gauge groups, including $SU(N)$, $SO(N)$, and $Sp(2N)$. By incorporating a group-theoretic dipole tagging scheme and an exact massive $2\to 3$ kinematic reconstruction module, our simulation precisely models mass-dependent phenomena such as the dead-cone effect, without resorting to the massless approximations frequently used in standard tools.
To decode the multidimensional branching histories, we mapped the jet constituents onto the Lund Jet Plane and introduced the Neural Sorter Mamba Network, an architecture designed to effectively capture the structural, causal correlations within the emission sequence. 

In the massless limit, we showed that distinct Casimir invariants and varying running couplings imprint unique hierarchical structures on the LJP. The NS-MambaNet exhibits high classification accuracy, rapidly building statistical significance to distinguish even highly degenerate gauge pairs (e.g., $SU(4)$ and $Sp(8)$) with a small sample of jet events.

The introduction of a finite dark gauge boson mass drastically alters the LJP topology, generating a distinct "soft radiation island" separated by a kinematic bifurcation. While heavier mediators dynamically shrink the allowed phase space and partially mask the purely group-theoretic differences governed by Casimir scaling, the network remains capable of disentangling the gauge structures given a moderately larger event sample.

Finally, by analyzing the classification performance under progressive infrared cutoffs on the $k_T$ variable, we demonstrated that the model's discrimination power is deeply rooted in the hard, perturbative branching history. 
While soft emissions are helpful for breaking specific leading-order Casimir degeneracies, the classifier exhibits exceptional robustness for the vast majority of symmetry pairs even when the non-perturbative regime is entirely filtered out.

The methodologies presented in this work outline a clear roadmap for the characterization phase post-discovery. 
We note that this study inherently assumes the constituents of dark jets are experimentally measurable, without specifying the exact phenomenological signature. In realistic scenarios, these constituents may decay into SM particles, yielding signals such as semi-visible jets, emerging jets, or flavor-specific lepton jets. Because the exact signature is highly model-dependent, its detailed modeling goes beyond the scope of this paper. 
Furthermore, our analysis focuses strictly on the discrimination between different dark gauge groups rather than the rejection of SM backgrounds. This approach is motivated by the fact that dark jets often exhibit macroscopic features, such as displaced vertices or unusual missing energy fractions, that inherently isolate them from standard QCD jets.
Regardless of the specific signature, by synergizing precision kinematics, first-principles representations like the LJP, and state-of-the-art deep learning architectures, future experimental programs can go beyond merely establishing the existence of a dark sector. Instead, they will be equipped to invert the perturbative footprints of these jets, reconstruct the ultraviolet nature of the dark interaction, and ultimately decode the fundamental symmetries governing dark matter.

\begin{acknowledgments}
This work was supported by the Natural Science Foundation of Sichuan Province under grant No. 2026NSFSC0034, by the National Natural Science Foundation of China (Project No. 11905149, 12505121), by the Joint Fund of Henan Province Science and Technology R$\&$D Program (Project No. 245200810077), by the Startup Research Fund of Henan Academy of Sciences (Project No. 20251820001), and by the Scientific and Technological Research Project of Henan Academy of Sciences (Project No. 20262320001). 
\end{acknowledgments}

\begin{appendices}
\section*{Appendices}
\section{Phase Space Limits} \label{sec:app1}

For the $2 \to 3$ splitting process $a + r \to b + c + r^\prime$ to be kinematically permitted, the following phase space conditions must be satisfied:
\begin{equation}
\label{eq:cc}
C_{min} \equiv \max(p_{T_{min}}^2,\ m_a^2,\ (m_b+m_c)^2) \leq m_{bc}^2 \leq s_{ar}+m_r^2-2\sqrt{s_{ar}(t+m_r^2)} \equiv C_{max}(t), 
\end{equation}
where $t$ represents the evolution variable, $p_{T_{min}}^2$ denotes the infrared cutoff, and $s_{ar}=(p_a+p_r)^2$ is the invariant mass squared of the dipole. We can derive the upper bound $C_{max}(t)$ by imposing energy-momentum conservation and the on-shell condition for the recoiling spectator particle $r$. The explicit derivation is outlined below.

In the center-of-mass frame of the $(a,\ r)$ dipole, the kinematics are given by:
\begin{equation}
\begin{split}
p_a+p_r &= p_b+p_c+p_{r^\prime} = 0\ , \\
E_{bc} &= \sqrt{p^2+m_{bc}^2}\ , \\
E_{r^\prime} &= \sqrt{p^2+m_{r^\prime}^2}\ , \\
s_{ar} &= (E_{bc} + E_{r^\prime})^2\ ,
\end{split}
\end{equation}
where $p = |p_{bc}| = |p_{r^\prime}|$. The center-of-mass energy squared $s_{ar}$ can be expanded and rearranged to yield:
\begin{equation}
m_{bc}^2 = s_{ar} + m_{r^\prime}^2 - 2\sqrt{s_{ar}(p^2+m_{r^\prime}^2)} \ .
\end{equation}
In our framework, the evolution variable $t$ is defined as $t = p_T^2 = z(1-z)m_{a^\prime}^2-(1-z)m_b^2-zm_c^2$. This corresponds to the squared transverse momentum of the emitted particle $b$ to the emitter $a^\prime$'s direction. Consequently, $t$ is bounded by the available three-momentum squared $p^2$, such that $t \leq p^2$. Given that the recoiled spectator $r^\prime$ remains on-shell, the inequality becomes
\begin{equation}
m_{bc}^2 \leq s_{ar} + m_r^2 - 2\sqrt{s_{ar}(t+m_r^2)}\ ,
\end{equation}
which precisely matches the definition of $C_{max}(t)$. Combining this with Eq.\ref{eq:cc} directly yields a fundamental kinematic constraint on $t$:
\begin{equation}
0 \leq t \leq \frac{(s_{ar}+C_{min}-m_r^2)^2}{4s_{ar}}-C_{min}\ .
\end{equation}

To determine the kinematically allowed phase space, we solve the two inequalities for $m_{bc}^2$. It is important to note that $m_{bc}^2$ acts as a function of both $t$ and $z$, parameterized as $m_{bc}^2(t, z)$. 
Our objective is to determine the allowed domain of $z$ for a given $t$. Evaluating the left-hand inequality first, a direct solution yields:
\begin{equation}
z_{1,\pm} = \frac{m_b^2+C_{min}-m_c^2}{2C_{min}} \pm \sqrt{(\frac{m_b^2+C_{min}-m_c^2}{2C_{min}})^2-\frac{t+m_b^2}{C_{min}}}\ .
\end{equation}
The nature of allowed region depends on the discriminant, $\Delta_1 = ((m_b^2+C_{min}-m_c^2)/(2C_{min}))^2-(t+m_b^2)/C_{min}$. 
Enforcing the physical requirement that $z$ must lie within the interval $[0, 1]$, we distinguish two scenarios
\begin{itemize}
\item $\Delta_1 \geq 0$, the solution set is $z \in [0, z_{1,-}(t)] \cup [z_{1,+}(t), 1]$.
\item $\Delta_1 < 0$, the inequality holds for all real values of $z \in [0, 1]$.
\end{itemize}
This solution implies that, for a suitable choice of massive parameters, the allowed range for $z$ splits into two disjoint intervals, creating a forbidden region between them. 

While the left-hand inequality does not restrict the range of $t$ (as a non-empty physical solution set for $z$ is guaranteed for any $t \geq 0$ regardless of the sign of $\Delta_1$), we must also satisfy the right-hand inequality. Solving it directly yields:
\begin{equation}
z_{2,\pm} = \frac{m_b^2+C_{max}(t)-m_c^2}{2C_{max}(t)} \pm \sqrt{(\frac{m_b^2+C_{max}(t)-m_c^2}{2C_{max}(t)})^2-\frac{t+m_b^2}{C_{max}(t)}}\ .
\end{equation}
Similarly, the valid solution set is determined by the sign of the discriminant $\Delta_2 = ((m_b^2+C_{max}(t)-m_c^2)/(2C_{max}(t)))^2-(t+m_b^2)/C_{max}(t)$:
\begin{itemize}
\item $\Delta_2 \geq 0$, the solution set is $z \in [z_{2,-}(t), z_{2,+}(t)]$.
\item $\Delta_2 < 0$, the inequality has no real solution.
\end{itemize}

As shown, the right-hand inequality introduces a new constraint on $t$, namely that a real solution exists only if $\Delta_2 \geq 0$. This condition mathematically reduces to solving a cubic inequality. while a general cubic equation may possess complex roots, the physical parameters of this system guarantee three distinct real root. Therefore, we can directly apply the trigonometric method to find the analytical solutions. 
Because converting the inequality to a standard cubic form may introduce extraneous roots, a straightforward consistency check is employed: each candidate root $(t_1, t_2, t_3)$ is substituted back into the original condition $\Delta_2 \geq 0$ to verify its physical validity.

\begin{figure}[t]
\includegraphics[width=0.45\textwidth]{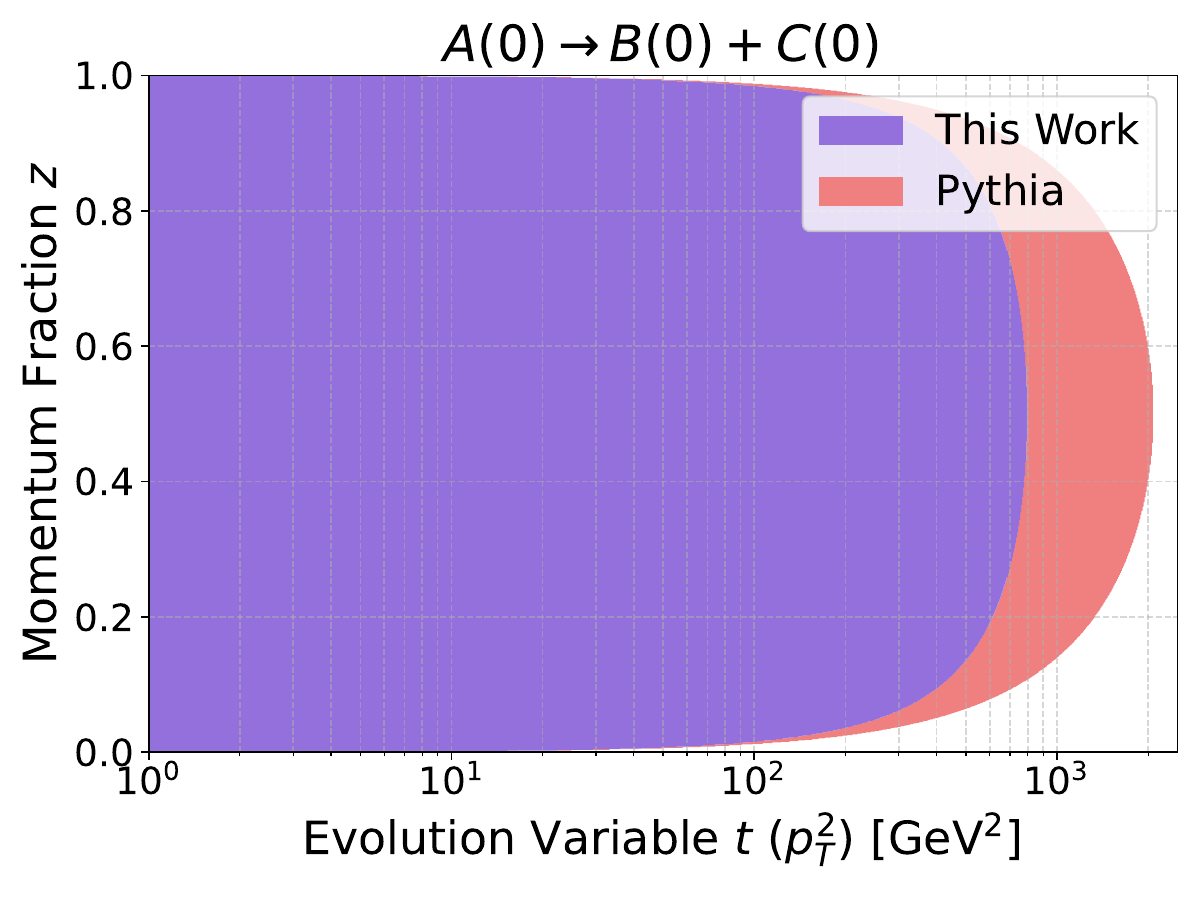}
\includegraphics[width=0.45\textwidth]{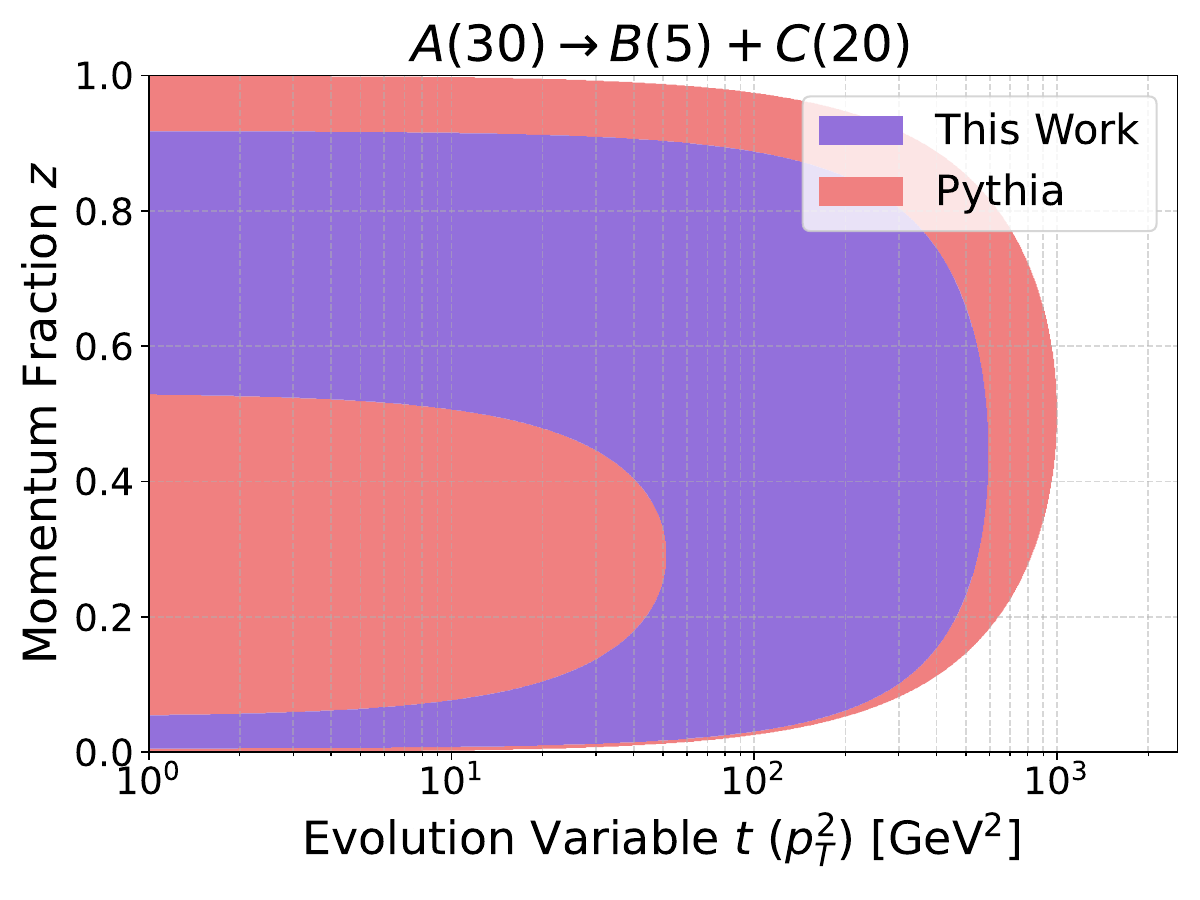}
\caption{Kinematically allowed phase space boundaries for $1 \to 2$ branching processes. 
The left panel displays the continuous, uniformly bounded phase space for a completely massless splitting, $A(0) \to B(0)+C(0)$. The right panel showcases a massive splitting scenario, $A(30) \to B(5)+C(20)$. The values in parentheses denote the respective particle masses. The red regions represent the corresponding phase space boundaries implemented in Pythia8.
\label{fig:limit}}
\end{figure}

In summary, by synthesizing all the constraints derived above, we establish the complete, kinematically allowed phase space for $z$ and $t$:
\begin{equation}
\begin{gathered}
z \in \begin{cases} [0,\ 1],\ \Delta_1 < 0 \\ [0,\ z_{1,-}(t)] \cup [z_{1,+}(t),\ 1],\ \Delta_1 \geq 0 \end{cases} \cap \ \ \ \begin{cases} \varnothing,\ \Delta_2 < 0 \\ [z_{2,-}(t),\ z_{2,+}(t)],\ \Delta_2 \geq 0 \end{cases}\ , \\
t \in [0,\ \frac{(s_{ar}+C_{min}-m_r^2)^2}{4s_{ar}}-C_{min}] \cap \begin{cases} [0,\ +\infty],\ \text{0 root for $\Delta_2 = 0$} \\ [t_1,\ +\infty],\ \text{1 root for $\Delta_2 = 0$} \\ [0,\ t_1] \cup [t_2,\ +\infty],\ \text{2 roots for $\Delta_2 = 0$} \\ [t_1,\ t_2] \cup [t_3,\ +\infty],\ \text{3 roots for $\Delta_2 = 0$} \end{cases}\ .
\end{gathered}
\end{equation}
The monotonic behavior of $\Delta z_1$ (increasing with $t$) and $\Delta z_2$ (decreasing with $t$) simplifies the numerical determination of the allowed range for $z$ over any given interval $[t_{min}, t_{max}]$.
The resulting phase space topologies are explicitly visualized in Figure~\ref{fig:limit}, which illustrates the features of continuous ranges and the disjoint $z$ boundaries that emerge under varying mass configurations, in contrast to the Pythia cases.

\section{Comparison with Pythia8} \label{sec:app2}

To highlight the distinctions between our algorithm and Pythia8, we further compare the first-splitting angular and the final state transverse momentum distributions.

\begin{figure}[htbp] 
\centering
\includegraphics[width=0.48\textwidth]{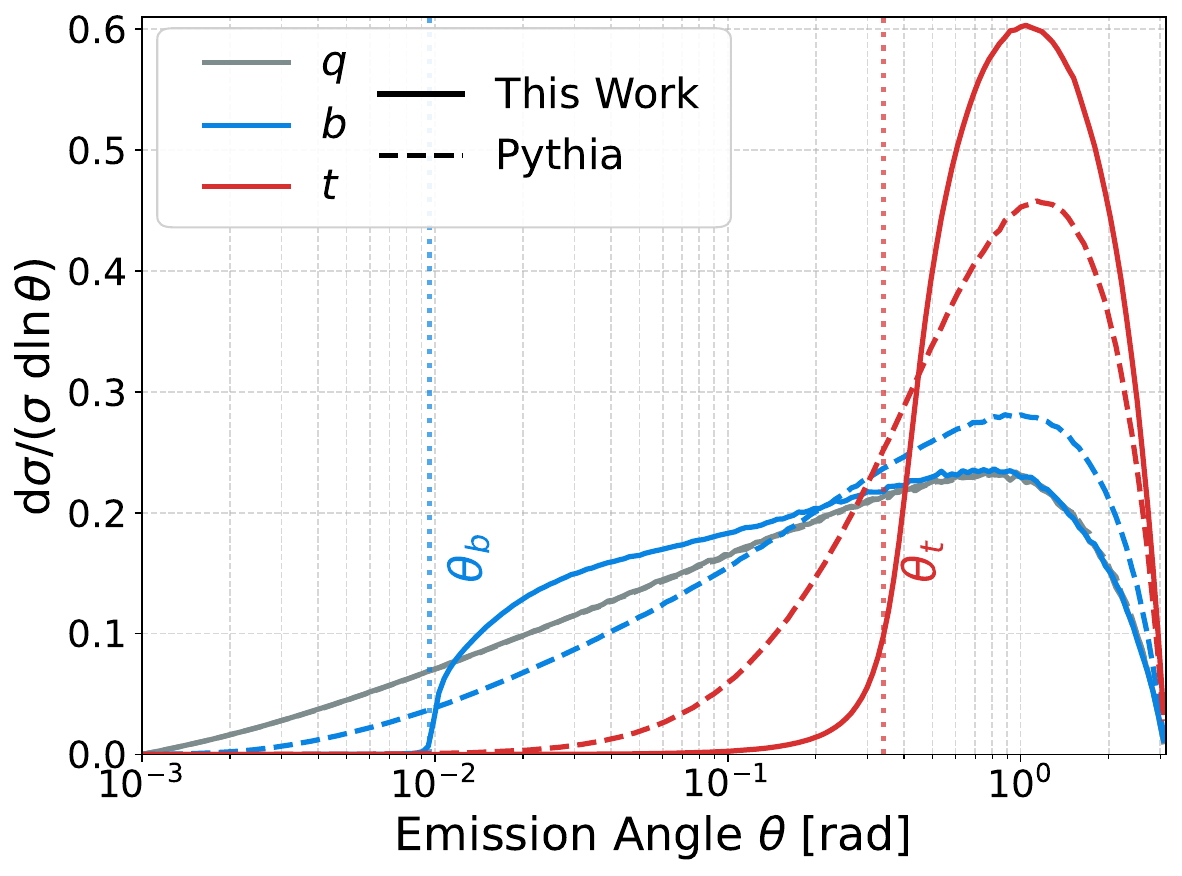}
\caption{The first-splitting angular distribution $\frac{1}{\sigma}\frac{d\sigma}{d\ln\theta}$ for light ($q$, grey), bottom ($b$, blue), and top ($t$, red) quarks. Solid lines represent this work; dashed lines represent Pythia8. Vertical dotted lines denote the physical dead-cone limits ($\theta_b$ and $\theta_t$).)
\label{fig:ps_dc}}
\end{figure}

Figure~\ref{fig:ps_dc} illustrates the first-splitting angular distribution for quarks with different masses. The single-emission kinematics are evaluated at a fixed dipole invariant mass of $\sqrt{s_{\text{dip}}} = 1$~TeV, with the shower cutoff scale set to $t_{\text{min}} = 0.25\text{ GeV}^2$. To generate the initial phase space, the evolution variable $t$ is sampled uniformly in logarithmic scale over $[\ln t_{\text{min}}, \ln(s_{\text{dip}}/4)]$, while the light-cone momentum fraction $z$ is sampled uniformly in $[10^{-3}, 1-10^{-3}]$. After applying the exact kinematic constraints and performing the massive reconstruction, each generated splitting is weighted by its corresponding splitting kernel.

In the massless limit ($q$), both algorithms yield identical distributions, confirming their mathematical consistency. For massive quarks ($b$ and $t$), both models successfully capture the dead-cone effect, showing a clear suppression of collinear emissions below the respective physical limits $\theta_b$ and $\theta_t$. However, our algorithm exhibits significantly stronger and more rapid suppression at small angles, while Pythia8 displays a more pronounced collinear tail. This difference arises because our framework strictly implements exact phase-space boundaries and full $2 \to 3$ kinematic reconstruction, while Pythia8 utilizes approximate boundaries combined with a multiplicative correction factor. This comparison confirms that our model achieves a more natural description of heavy particle radiation.

\begin{figure}[htbp] \centering
\includegraphics[width=0.45\textwidth]{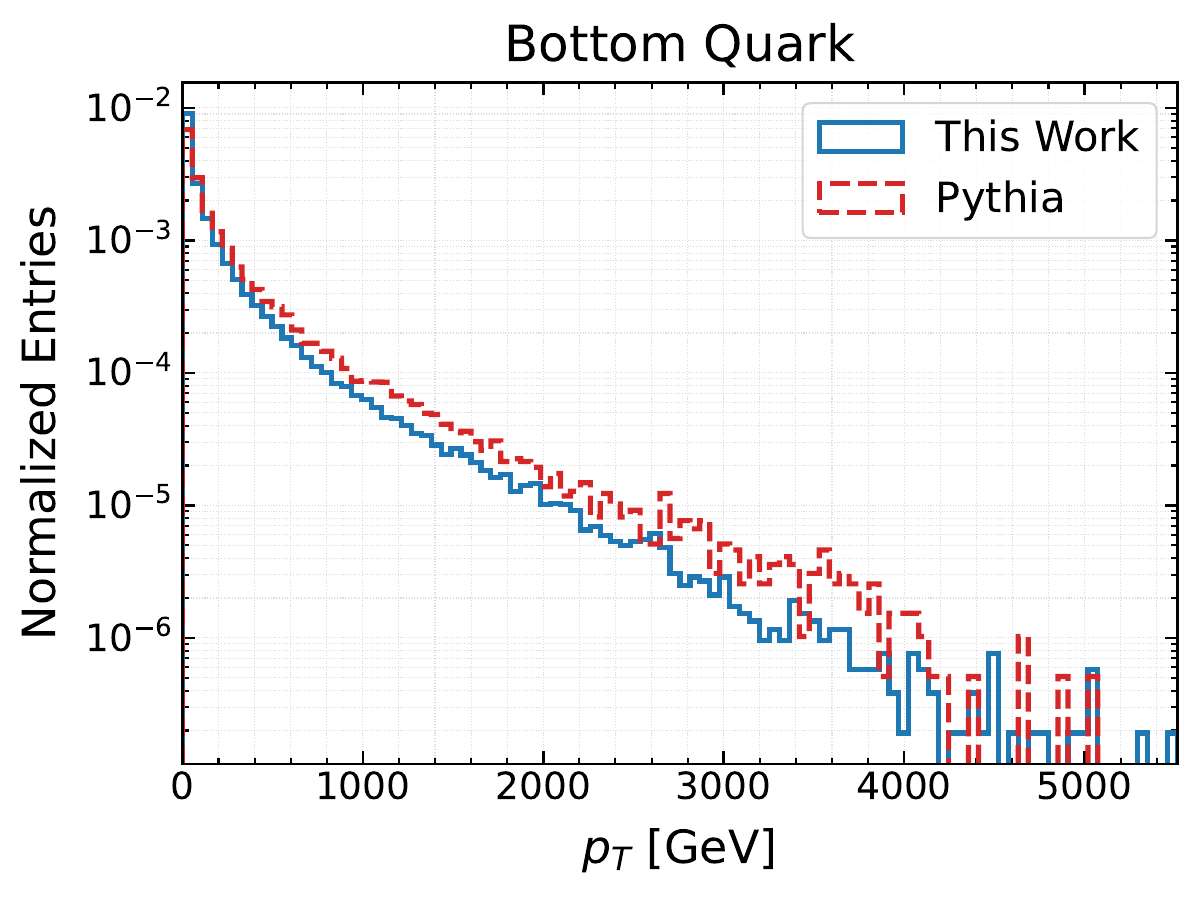}
\includegraphics[width=0.45\textwidth]{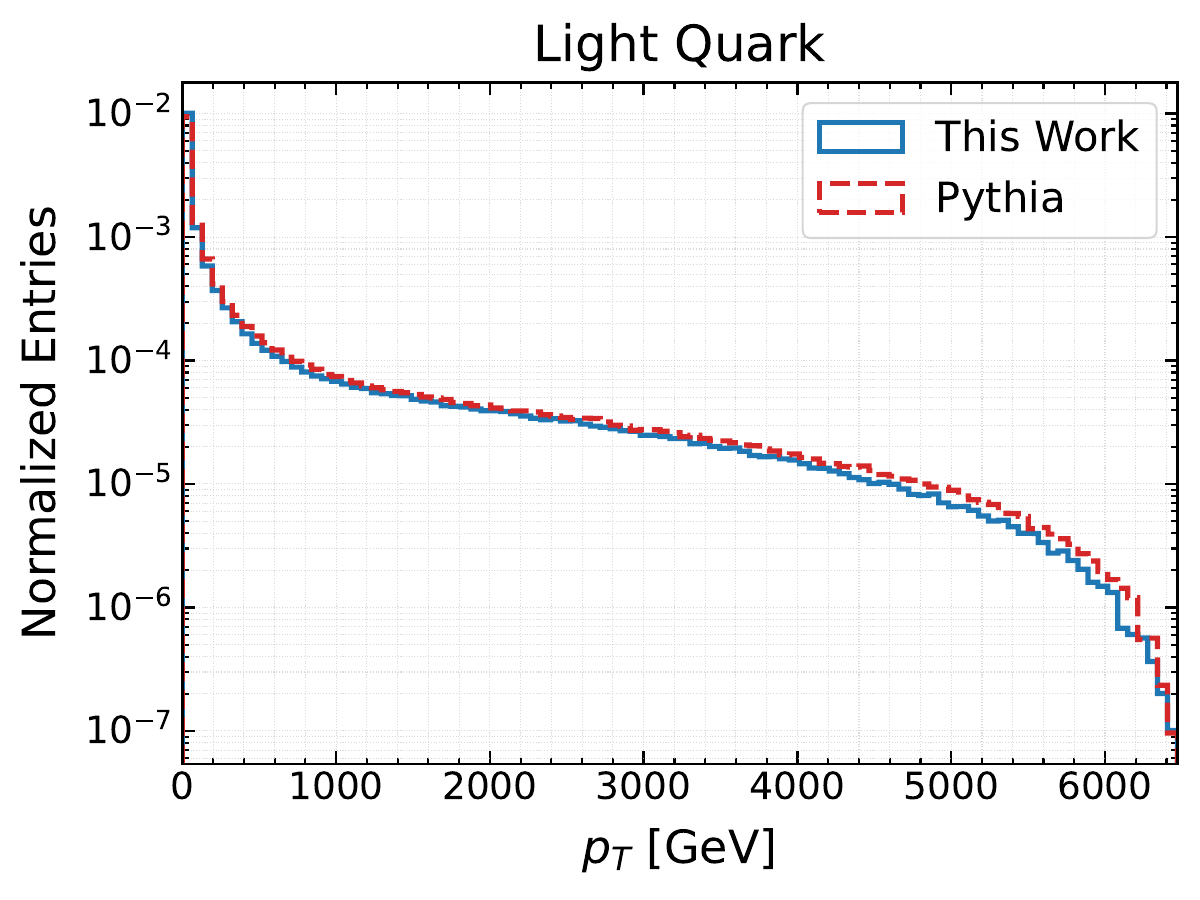} \\
\includegraphics[width=0.45\textwidth]{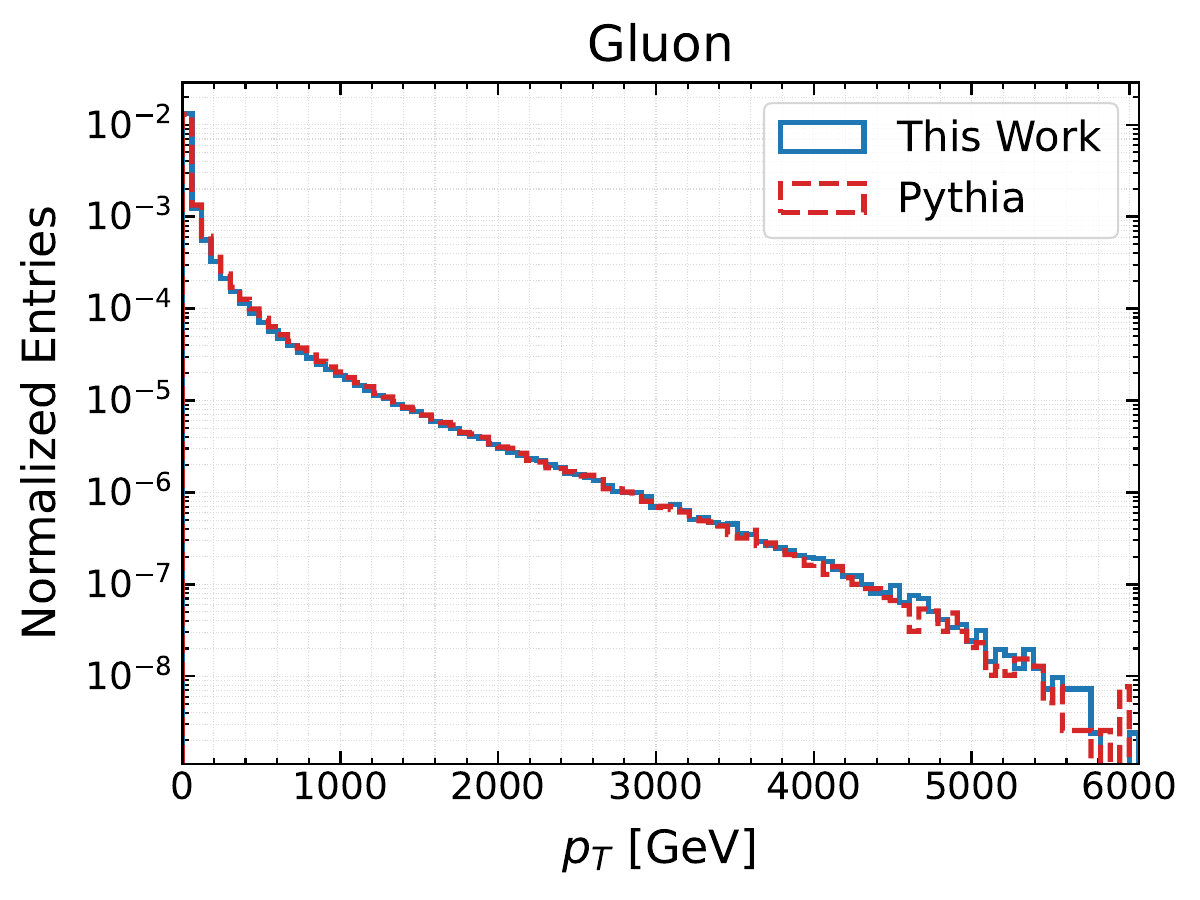}
\includegraphics[width=0.45\textwidth]{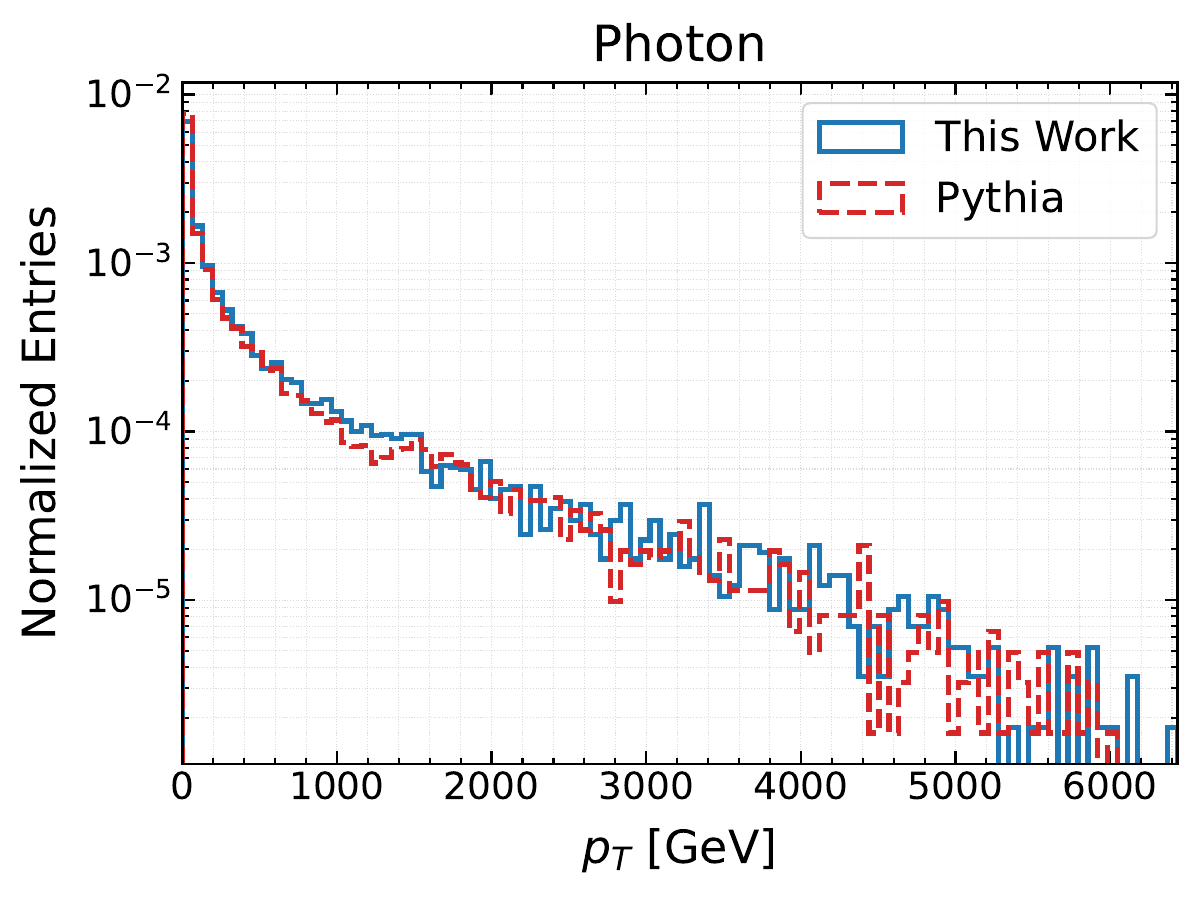}
\caption{The transverse momentum ($p_T$) distributions of final-state $b$ quarks (top left), light quarks (top right), gluons (bottom left), and photons (bottom right) in the full shower. Solid blue lines represent the results of this work, while dashed red lines represent Pythia8.
\label{fig:mom}}
\end{figure}

Figure~\ref{fig:mom} illustrates the final-state transverse momentum ($p_T$) distribution of different particle species prior to hadronization. To obtain these distributions, we simulate the full parton shower with $10^5$ events generated from $e^+ e^- \to Z \to q\bar{q}$ processes at a center-of-mass energy of $\sqrt{s} = 13$~TeV, where the $Z$ boson mass is set to $13$~TeV. The shower evolution is restricted to final-state radiation with the infrared cutoff scale set to $t_{\text{min}} = 0.25\text{ GeV}^2$, while initial-state radiation, multi-parton interactions, and hadronization are disabled in Pythia. 

For massless particles, including light quarks, gluons, and photons, the $p_T$ distributions predicted by our framework and Pythia8 are in excellent agreement across the entire kinematic range spanning several orders of magnitude. This consistent behavior verifies that our algorithm correctly reproduces the standard massless emission spectrum. For the massive $b$ quark, a slight discrepancy is observed in the intermediate to high $p_T$ region, where our algorithm predicts a slightly softer spectrum compared to Pythia8. This difference is directly related to our exact treatment of massive phase space boundaries and kinematic reconstruction, which naturally restricts the high $p_T$ emission phase space for heavy quarks.


\end{appendices}

\bibliographystyle{jhep}
\bibliography{jhep}

\end{document}